June 2009

# Studies of Cosmic Ray Composition and Air Shower Structure with the Pierre Auger Observatory

Presentations for the
31st International Cosmic Ray Conference, Łódź , Poland, July 2009

# PIERRE AUGER COLLABORATION

**J. Abraham[8], P. Abreu[71], M. Aglietta[54], C. Aguirre[12], E.J. Ahn[87], D. Allard[31], I. Allekotte[1], J. Allen[90], J. Alvarez-Muñiz[78], M. Ambrosio[48], L. Anchordoqui[104], S. Andringa[71], A. Anzalone[53], C. Aramo[48], E. Arganda[75], S. Argirò[51], K. Arisaka[95], F. Arneodo[55], F. Arqueros[75], T. Asch[38], H. Asorey[1], P. Assis[71], J. Aublin[33], M. Ave[96], G. Avila[10], T. Bäcker[42], D. Badagnani[6], K.B. Barber[11], A.F. Barbosa[14], S.L.C. Barroso[20], B. Baughman[92], P. Bauleo[85], J.J. Beatty[92], T. Beau[31], B.R. Becker[101], K.H. Becker[36], A. Bellétoile[34], J.A. Bellido[11, 93], S. BenZvi[103], C. Berat[34], P. Bernardini[47], X. Bertou[1], P.L. Biermann[39], P. Billoir[33], O. Blanch-Bigas[33], F. Blanco[75], C. Bleve[47], H. Blümer[41, 37], M. Boháčová[96, 27], D. Boncioli[49], C. Bonifazi[33], R. Bonino[54], N. Borodai[69], J. Brack[85], P. Brogueira[71], W.C. Brown[86], R. Bruijn[81], P. Buchholz[42], A. Bueno[77], R.E. Burton[83], N.G. Busca[31], K.S. Caballero-Mora[41], L. Caramete[39], R. Caruso[50], W. Carvalho[17], A. Castellina[54], O. Catalano[53], L. Cazon[96], R. Cester[51], J. Chauvin[34], A. Chiavassa[54], J.A. Chinellato[18], A. Chou[87, 90], J. Chudoba[27], J. Chye[89d], R.W. Clay[11], E. Colombo[2], R. Conceição[71], B. Connolly[102], F. Contreras[9], J. Coppens[65, 67], A. Cordier[32], U. Cotti[63], S. Coutu[93], C.E. Covault[83], A. Creusot[73], A. Criss[93], J. Cronin[96], A. Curutiu[39], S. Dagoret-Campagne[32], R. Dallier[35], K. Daumiller[37], B.R. Dawson[11], R.M. de Almeida[18], M. De Domenico[50], C. De Donato[46], S.J. de Jong[65], G. De La Vega[8], W.J.M. de Mello Junior[18], J.R.T. de Mello Neto[23], I. De Mitri[47], V. de Souza[16], K.D. de Vries[66], G. Decerprit[31], L. del Peral[76], O. Deligny[30], A. Della Selva[48], C. Delle Fratte[49], H. Dembinski[40], C. Di Giulio[49], J.C. Diaz[89], P.N. Diep[105], C. Dobrigkeit [18], J.C. D'Olivo[64], P.N. Dong[105], A. Dorofeev[88], J.C. dos Anjos[14], M.T. Dova[6], D. D'Urso[48], I. Dutan[39], M.A. DuVernois[98], R. Engel[37], M. Erdmann[40], C.O. Escobar[18], A. Etchegoyen[2], P. Facal San Luis[96, 78], H. Falcke[65, 68], G. Farrar[90], A.C. Fauth[18], N. Fazzini[87], F. Ferrer[83], A. Ferrero[2], B. Fick[89], A. Filevich[2], A. Filipčič[72, 73], I. Fleck[42], S. Fliescher[40], C.E. Fracchiolla[85], E.D. Fraenkel[66], W. Fulgione[54], R.F. Gamarra[2], S. Gambetta[44], B. García[8], D. García Gámez[77], D. Garcia-Pinto[75], X. Garrido[37, 32], G. Gelmini[95], H. Gemmeke[38], P.L. Ghia[30, 54], U. Giaccari[47], M. Giller[70], H. Glass[87], L.M. Goggin[104], M.S. Gold[101], G. Golup[1], F. Gomez Albarracin[6], M. Gómez Berisso[1], P. Gonçalves[71], M. Gonçalves do Amaral[24], D. Gonzalez[41], J.G. Gonzalez[77, 88], D. Góra[41, 69], A. Gorgi[54], P. Gouffon[17], S.R. Gozzini[81], E. Grashorn[92], S. Grebe[65], M. Grigat[40], A.F. Grillo[55], Y. Guardincerri[4], F. Guarino[48], G.P. Guedes[19], J. Gutiérrez[76], J.D. Hague[101], V. Halenka[28], P. Hansen[6], D. Harari[1], S. Harmsma[66, 67], J.L. Harton[85], A. Haungs[37], M.D. Healy[95], T. Hebbeker[40], G. Hebrero[76], D. Heck[37], C. Hojvat[87], V.C. Holmes[11], P. Homola[69], J.R. Hörandel[65], A. Horneffer[65], M. Hrabovský[28, 27], T. Huege[37], M. Hussain[73], M. Iarlori[45], A. Insolia[50], F. Ionita[96], A. Italiano[50], S. Jiraskova[65], M. Kaducak[87], K.H. Kampert[36], T. Karova[27], P. Kasper[87], B. Kégl[32], B. Keilhauer[37], E. Kemp[18], R.M. Kieckhafer[89], H.O. Klages[37], M. Kleifges[38], J. Kleinfeller[37], R. Knapik[85], J. Knapp[81], D.-H. Koang[34], A. Krieger[2], O. Krömer[38], D. Kruppke-Hansen[36], F. Kuehn[87], D. Kuempel[36], N. Kunka[38], A. Kusenko[95], G. La Rosa[53], C. Lachaud[31], B.L. Lago[23], P. Lautridou[35], M.S.A.B. Leão[22], D. Lebrun[34], P. Lebrun[87], J. Lee[95], M.A. Leigui de Oliveira[22], A. Lemiere[30], A. Letessier-Selvon[33], M. Leuthold[40], I. Lhenry-Yvon[30], R. López[59], A. Lopez Agüera[78], K. Louedec[32], J. Lozano Bahilo[77], A. Lucero[54], H. Lyberis[30], M.C. Maccarone[53], C. Macolino[45], S. Maldera[54], D. Mandat[27], P. Mantsch[87], A.G. Mariazzi[6], I.C. Maris[41], H.R. Marquez Falcon[63], D. Martello[47], O. Martínez Bravo[59], H.J. Mathes[37], J. Matthews[88, 94], J.A.J. Matthews[101], G. Matthiae[49], D. Maurizio[51], P.O. Mazur[87], M. McEwen[76], R.R. McNeil[88], G. Medina-Tanco[64], M. Melissas[41], D. Melo[51], E. Menichetti[51], A. Menshikov[38], R. Meyhandan[14], M.I. Micheletti[2], G. Miele[48], W. Miller[101], L. Miramonti[46], S. Mollerach[1], M. Monasor[75], D. Monnier Ragaigne[32], F. Montanet[34], B. Morales[64], C. Morello[54], J.C. Moreno[6], C. Morris[92], M. Mostafá[85], C.A. Moura[48], S. Mueller[37], M.A. Muller[18], R. Mussa[51], G. Navarra[54], J.L. Navarro[77], S. Navas[77], P. Necesal[27], L. Nellen[64], C. Newman-Holmes[87], D. Newton[81], P.T. Nhung[105], N. Nierstenhoefer[36], D. Nitz[89], D. Nosek[26], L. Nožka[27], M. Nyklicek[27], J. Oehlschläger[37], A. Olinto[96], P. Oliva[36], V.M. Olmos-Gilbaja[78], M. Ortiz[75], N. Pacheco[76], D. Pakk Selmi-Dei[18], M. Palatka[27], J. Pallotta[3], G. Parente[78], E. Parizot[31], S. Parlati[55], S. Pastor[74], M. Patel[81], T. Paul[91], V. Pavlidou[96c], K. Payet[34], M. Pech[27], J. Pękala[69], I.M. Pepe[21], L. Perrone[52], R. Pesce[44], E. Petermann[100], S. Petrera[45], P. Petrinca[49], A. Petrolini[44], Y. Petrov[85], J. Petrovic[67], C. Pfendner[103], R. Piegaia[4], T. Pierog[37], M. Pimenta[71], T. Pinto[74], V. Pirronello[50], O. Pisanti[48], M. Platino[2], J. Pochon[1], V.H. Ponce[1], M. Pontz[42], P. Privitera[96], M. Prouza[27], E.J. Quel[3], J. Rautenberg[36], O. Ravel[35],**

D. Ravignani[2], A. Redondo[76], B. Revenu[35], F.A.S. Rezende[14], J. Ridky[27], S. Riggi[50], M. Risse[36], C. Rivière[34], V. Rizi[45], C. Robledo[59], G. Rodriguez[49], J. Rodriguez Martino[50], J. Rodriguez Rojo[9], I. Rodriguez-Cabo[78], M.D. Rodríguez-Frías[76], G. Ros[75, 76], J. Rosado[75], T. Rossler[28], M. Roth[37], B. Rouillé-d'Orfeuil[31], E. Roulet[1], A.C. Rovero[7], F. Salamida[45], H. Salazar[59b], G. Salina[49], F. Sánchez[64], M. Santander[9], C.E. Santo[71], E.M. Santos[23], F. Sarazin[84], S. Sarkar[79], R. Sato[9], N. Scharf[40], V. Scherini[36], H. Schieler[37], P. Schiffer[40], A. Schmidt[38], F. Schmidt[96], T. Schmidt[41], O. Scholten[66], H. Schoorlemmer[65], J. Schovancova[27], P. Schovánek[27], F. Schroeder[37], S. Schulte[40], F. Schüssler[37], D. Schuster[84], S.J. Sciutto[6], M. Scuderi[50], A. Segreto[53], D. Semikoz[31], M. Settimo[47], R.C. Shellard[14, 15], I. Sidelnik[2], B.B. Siffert[23], A. Śmiałkowski[70], R. Šmída[27], B.E. Smith[81], G.R. Snow[100], P. Sommers[93], J. Sorokin[11], H. Spinka[82, 87], R. Squartini[9], E. Strazzeri[32], A. Stutz[34], F. Suarez[2], T. Suomijärvi[30], A.D. Supanitsky[64], M.S. Sutherland[92], J. Swain[91], Z. Szadkowski[70], A. Tamashiro[7], A. Tamburro[41], T. Tarutina[6], O. Taşcău[36], R. Tcaciuc[42], D. Tcherniakhovski[38], D. Tegolo[58], N.T. Thao[105], D. Thomas[85], R. Ticona[13], J. Tiffenberg[4], C. Timmermans[67, 65], W. Tkaczyk[70], C.J. Todero Peixoto[22], B. Tomé[71], A. Tonachini[51], I. Torres[59], P. Travnicek[27], D.B. Tridapalli[17], G. Tristram[31], E. Trovato[50], M. Tueros[6], R. Ulrich[37], M. Unger[37], M. Urban[32], J.F. Valdés Galicia[64], I. Valiño[37], L. Valore[48], A.M. van den Berg[66], J.R. Vázquez[75], R.A. Vázquez[78], D. Veberič[73, 72], A. Velarde[13], T. Venters[96], V. Verzi[49], M. Videla[8], L. Villaseñor[63], S. Vorobiov[73], L. Voyvodic[87‡], H. Wahlberg[6], P. Wahrlich[11], O. Wainberg[2], D. Warner[85], A.A. Watson[81], S. Westerhoff[103], B.J. Whelan[11], G. Wieczorek[70], L. Wiencke[84], B. Wilczyńska[69], H. Wilczyński[69], C. Wileman[81], M.G. Winnick[11], H. Wu[32], B. Wundheiler[2], T. Yamamoto[96a], P. Younk[85], G. Yuan[88], A. Yushkov[48], E. Zas[78], D. Zavrtanik[73, 72], M. Zavrtanik[72, 73], I. Zaw[90], A. Zepeda[60b], M. Ziolkowski[42]

[1] *Centro Atómico Bariloche and Instituto Balseiro (CNEA-UNCuyo-CONICET), San Carlos de Bariloche, Argentina*
[2] *Centro Atómico Constituyentes (Comisión Nacional de Energía Atómica/CONICET/UTN- FRBA), Buenos Aires, Argentina*
[3] *Centro de Investigaciones en Láseres y Aplicaciones, CITEFA and CONICET, Argentina*
[4] *Departamento de Física, FCEyN, Universidad de Buenos Aires y CONICET, Argentina*
[6] *IFLP, Universidad Nacional de La Plata and CONICET, La Plata, Argentina*
[7] *Instituto de Astronomía y Física del Espacio (CONICET), Buenos Aires, Argentina*
[8] *National Technological University, Faculty Mendoza (CONICET/CNEA), Mendoza, Argentina*
[9] *Pierre Auger Southern Observatory, Malargüe, Argentina*
[10] *Pierre Auger Southern Observatory and Comisión Nacional de Energía Atómica, Malargüe, Argentina*
[11] *University of Adelaide, Adelaide, S.A., Australia*
[12] *Universidad Catolica de Bolivia, La Paz, Bolivia*
[13] *Universidad Mayor de San Andrés, Bolivia*
[14] *Centro Brasileiro de Pesquisas Fisicas, Rio de Janeiro, RJ, Brazil*
[15] *Pontifícia Universidade Católica, Rio de Janeiro, RJ, Brazil*
[16] *Universidade de São Paulo, Instituto de Física, São Carlos, SP, Brazil*
[17] *Universidade de São Paulo, Instituto de Física, São Paulo, SP, Brazil*
[18] *Universidade Estadual de Campinas, IFGW, Campinas, SP, Brazil*
[19] *Universidade Estadual de Feira de Santana, Brazil*
[20] *Universidade Estadual do Sudoeste da Bahia, Vitoria da Conquista, BA, Brazil*
[21] *Universidade Federal da Bahia, Salvador, BA, Brazil*
[22] *Universidade Federal do ABC, Santo André, SP, Brazil*
[23] *Universidade Federal do Rio de Janeiro, Instituto de Física, Rio de Janeiro, RJ, Brazil*
[24] *Universidade Federal Fluminense, Instituto de Fisica, Niterói, RJ, Brazil*
[26] *Charles University, Faculty of Mathematics and Physics, Institute of Particle and Nuclear Physics, Prague, Czech Republic*
[27] *Institute of Physics of the Academy of Sciences of the Czech Republic, Prague, Czech Republic*
[28] *Palacký University, Olomouc, Czech Republic*
[30] *Institut de Physique Nucléaire d'Orsay (IPNO), Université Paris 11, CNRS-IN2P3, Orsay, France*
[31] *Laboratoire AstroParticule et Cosmologie (APC), Université Paris 7, CNRS-IN2P3, Paris, France*
[32] *Laboratoire de l'Accélérateur Linéaire (LAL), Université Paris 11, CNRS-IN2P3, Orsay, France*
[33] *Laboratoire de Physique Nucléaire et de Hautes Energies (LPNHE), Universités Paris 6 et Paris 7, Paris Cedex 05, France*

[34] Laboratoire de Physique Subatomique et de Cosmologie (LPSC), Université Joseph Fourier, INPG, CNRS-IN2P3, Grenoble, France
[35] SUBATECH, Nantes, France
[36] Bergische Universität Wuppertal, Wuppertal, Germany
[37] Forschungszentrum Karlsruhe, Institut für Kernphysik, Karlsruhe, Germany
[38] Forschungszentrum Karlsruhe, Institut für Prozessdatenverarbeitung und Elektronik, Karlsruhe, Germany
[39] Max-Planck-Institut für Radioastronomie, Bonn, Germany
[40] RWTH Aachen University, III. Physikalisches Institut A, Aachen, Germany
[41] Universität Karlsruhe (TH), Institut für Experimentelle Kernphysik (IEKP), Karlsruhe, Germany
[42] Universität Siegen, Siegen, Germany
[44] Dipartimento di Fisica dell'Università and INFN, Genova, Italy
[45] Università dell'Aquila and INFN, L'Aquila, Italy
[46] Università di Milano and Sezione INFN, Milan, Italy
[47] Dipartimento di Fisica dell'Università del Salento and Sezione INFN, Lecce, Italy
[48] Università di Napoli "Federico II" and Sezione INFN, Napoli, Italy
[49] Università di Roma II "Tor Vergata" and Sezione INFN, Roma, Italy
[50] Università di Catania and Sezione INFN, Catania, Italy
[51] Università di Torino and Sezione INFN, Torino, Italy
[52] Dipartimento di Ingegneria dell'Innovazione dell'Università del Salento and Sezione INFN, Lecce, Italy
[53] Istituto di Astrofisica Spaziale e Fisica Cosmica di Palermo (INAF), Palermo, Italy
[54] Istituto di Fisica dello Spazio Interplanetario (INAF), Università di Torino and Sezione INFN, Torino, Italy
[55] INFN, Laboratori Nazionali del Gran Sasso, Assergi (L'Aquila), Italy
[58] Università di Palermo and Sezione INFN, Catania, Italy
[59] Benemérita Universidad Autónoma de Puebla, Puebla, Mexico
[60] Centro de Investigación y de Estudios Avanzados del IPN (CINVESTAV), México, D.F., Mexico
[61] Instituto Nacional de Astrofisica, Optica y Electronica, Tonantzintla, Puebla, Mexico
[63] Universidad Michoacana de San Nicolas de Hidalgo, Morelia, Michoacan, Mexico
[64] Universidad Nacional Autonoma de Mexico, Mexico, D.F., Mexico
[65] IMAPP, Radboud University, Nijmegen, Netherlands
[66] Kernfysisch Versneller Instituut, University of Groningen, Groningen, Netherlands
[67] NIKHEF, Amsterdam, Netherlands
[68] ASTRON, Dwingeloo, Netherlands
[69] Institute of Nuclear Physics PAN, Krakow, Poland
[70] University of Łódź, Łódź, Poland
[71] LIP and Instituto Superior Técnico, Lisboa, Portugal
[72] J. Stefan Institute, Ljubljana, Slovenia
[73] Laboratory for Astroparticle Physics, University of Nova Gorica, Slovenia
[74] Instituto de Física Corpuscular, CSIC-Universitat de València, Valencia, Spain
[75] Universidad Complutense de Madrid, Madrid, Spain
[76] Universidad de Alcalá, Alcalá de Henares (Madrid), Spain
[77] Universidad de Granada & C.A.F.P.E., Granada, Spain
[78] Universidad de Santiago de Compostela, Spain
[79] Rudolf Peierls Centre for Theoretical Physics, University of Oxford, Oxford, United Kingdom
[81] School of Physics and Astronomy, University of Leeds, United Kingdom
[82] Argonne National Laboratory, Argonne, IL, USA
[83] Case Western Reserve University, Cleveland, OH, USA
[84] Colorado School of Mines, Golden, CO, USA
[85] Colorado State University, Fort Collins, CO, USA
[86] Colorado State University, Pueblo, CO, USA
[87] Fermilab, Batavia, IL, USA
[88] Louisiana State University, Baton Rouge, LA, USA
[89] Michigan Technological University, Houghton, MI, USA
[90] New York University, New York, NY, USA
[91] Northeastern University, Boston, MA, USA
[92] Ohio State University, Columbus, OH, USA
[93] Pennsylvania State University, University Park, PA, USA
[94] Southern University, Baton Rouge, LA, USA
[95] University of California, Los Angeles, CA, USA

[96] *University of Chicago, Enrico Fermi Institute, Chicago, IL, USA*
[98] *University of Hawaii, Honolulu, HI, USA*
[100] *University of Nebraska, Lincoln, NE, USA*
[101] *University of New Mexico, Albuquerque, NM, USA*
[102] *University of Pennsylvania, Philadelphia, PA, USA*
[103] *University of Wisconsin, Madison, WI, USA*
[104] *University of Wisconsin, Milwaukee, WI, USA*
[105] *Institute for Nuclear Science and Technology (INST), Hanoi, Vietnam*
‡ *Deceased*
[a] *at Konan University, Kobe, Japan*
[b] *On leave of absence at the Instituto Nacional de Astrofisica, Optica y Electronica*
[c] *at Caltech, Pasadena, USA*
[d] *at Hawaii Pacific University*

# Measurement of the average depth of shower maximum and its fluctuations with the Pierre Auger Observatory

**J. A. Bellido*** **for the Pierre Auger Collaboration**†

* *Physics Department, The University of Adelaide, S.A. - 5005, Australia*
† *Observatorio Pierre Auger, Av. San Martin Norte 304, 5613 Malargüe, Argentina*

***Abstract*. The atmospheric depth $X_{\max}$ where an air shower reaches its maximum size is measured shower-by-shower with a resolution of 20 g cm$^{-2}$ on average using the air fluorescence telescopes of the Pierre Auger Observatory. The mean value $\langle X_{\max}\rangle$ and the RMS width of the Xmax distribution will be reported for 13 different logarithmic energy intervals above 1 EeV.**

## I. Introduction

The Pierre Auger Observatory has recently taken steps towards unveiling the mysterious origin of the most energetic cosmic rays. In a recent publication we reported the measured energy spectrum [1], which has confirmed, with improved statistics, a suppression in the spectrum beyond about $10^{19.6}$ eV. This is consistent with the predicted GZK pion photoproduction or nuclear photodisintegration [2], but it could also be the result of the intrinsic source spectrum. Another important feature observed in the energy spectrum at energies between $10^{18}$ and $10^{19}$ eV is the so-called ankle or dip. It is suggested that a source transition from galactic to extragalactic is the cause of this feature [3], but it has also been suggested that the galactic-extragalactic transition happens at a lower energy and that at around $10^{18.5}$ eV cosmic rays are mainly extragalactic protons that interact with the CMB radiation producing the dip by $e^{\pm}$ pair production [4] (see [5] for a discussion of both models). Another Auger publication has shown evidence for an anisotropy in the arrival directions of the most energetic cosmic rays [6].

A great deal of information on the nature of the cosmic ray sources and the characteristics of the particle propagation is contained in the energy spectrum and in the observed anisotropy. Additional information on the cosmic ray mass composition can help to complete the overall picture.

The fluorescence detector (FD) of the Pierre Auger Observatory can be used to measure with good resolution the shower longitudinal profile and the depth at which the shower reaches its maximum size ($X_{\max}$). At a given energy, the average $X_{\max}$ and the width of the $X_{\max}$ distribution are both correlated with the cosmic ray mass composition [8]. Proton showers penetrate deeper into the atmosphere (larger values of $X_{\max}$) and have wider $X_{\max}$ distributions than heavy nuclei.

The mass composition interpretation of the measured mean and width of the $X_{\max}$ distribution depend on the assumed hadronic model. The problem is that at these high energies, the uncertainties on the predictions from the models are unknown because they are an extrapolation of the physics from lower energies.

## II. Data Analysis

We have used hybrid events to measure the longitudinal profiles of air showers. These are events observed simultaneously by the FD and by at least one surface detector. The information from the surface detector allows us to constrain the geometry of the air shower. This hybrid constraint on the geometry is not efficient when the time duration of the event as seen by the FD is small (less than 0.5 $\mu$s), because the time synchronization between the surface detectors and the FD is of the order of 0.1 $\mu$s. To exclude such short-duration events we have rejected showers with directions pointing towards the FD by requiring that the minimum viewing angle be greater than 20° (this cut also removes events with a large fraction of direct Cherenkov light). The hybrid reconstruction has an average angular resolution of 0.6° [10]. Good resolution in the reconstructed geometry is the first step towards good resolution in $X_{\max}$ measurements.

**Profile quality cuts:** Our aim is to measure $X_{\max}$ with an average resolution of 20 g cm$^{-2}$. To achieve this goal we have used Monte Carlo simulated data to design a set of quality cuts for the observed profiles. The reconstructed [7] $X_{\max}$ should lie within the observed shower profile, the length, in g cm$^{-2}$, of the observed profile should be at least 320 g cm$^{-2}$, and the reduced $\chi^2$ of a fit with a Gaisser-Hillas function [11] should not exceed 2.5. Moreover, shower profiles with insignificant curvature around the reconstructed $X_{\max}$ are rejected by requiring that the $\chi^2$ of a linear fit to the longitudinal profile exceeds the Gaisser-Hillas fit $\chi^2$ by at least four. Finally, the estimated uncertainties of the shower maximum and total energy must be smaller than 40 g cm$^{-2}$ and 20%, respectively.

To check the $X_{\max}$ resolution we have used stereo events. Stereo events have an average energy of $10^{19}$ eV. Figure 1 shows a comparison between the $X_{\max}$ values independently reconstructed with each FD. The factor $1/\sqrt{2}$ (in the x axis) is to take into account that the RMS of the $\Delta X_{\max}$ would correspond to the convolution of the two resolutions,

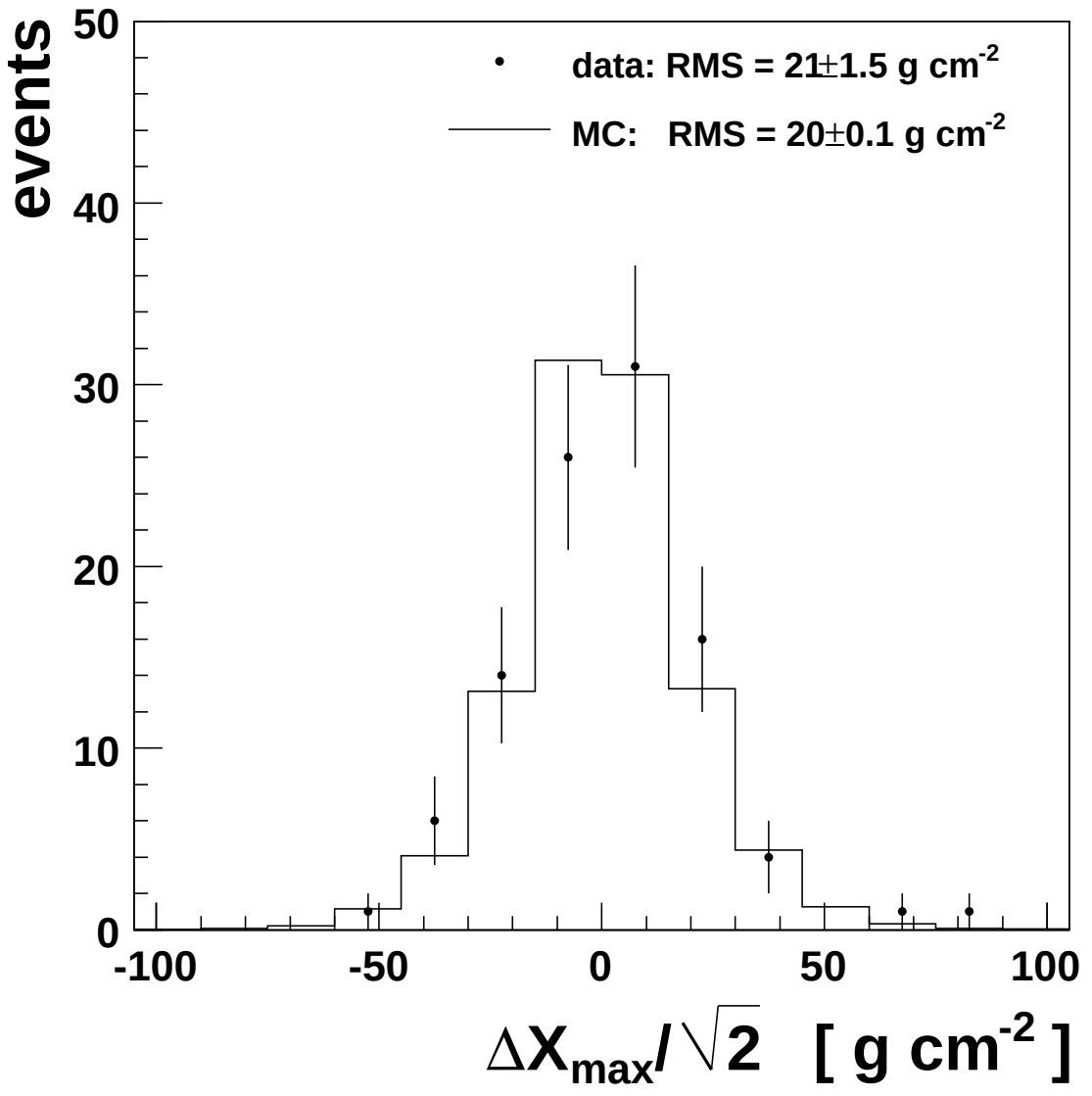

Fig. 1. Difference of $X_{\max}$ reconstructions for showers that have been observed by at least two FD sites (in real data and in MC).

$$\mathrm{RMS}(\Delta X_{\max})^2 = \sigma_1^2 + \sigma_2^2. \quad (1)$$

The two $X_{\max}$ resolutions, $\sigma_1$ and $\sigma_2$, are not necessarily the same, however we can approximate them to an average $X_{\max}$ resolution ($\sigma_1 \approx \sigma_2 \approx \sigma$) and rewrite equation 1,

$$\sigma = \mathrm{RMS}(\Delta X_{\max}/\sqrt{2}). \quad (2)$$

The average resolution ($\sigma$) for the reconstructed $X_{\max}$ in stereo events is $21 \pm 1.5\,\mathrm{g\,cm^{-2}}$ (obtained from figure 1). This resolution is consistent with the resolution obtained with stereo events generated with MC simulations (also shown in figure 1).

**Cuts for an unbiased measurement of the $X_{\max}$ distribution:** To ensure a trigger probability close to unity for protons and iron at energies above $10^{18}$ eV, we apply energy-dependent cuts on the zenith angle and the maximum distance from the shower core to the nearest surface detector. The Auger FD has a field of view ranging from 1.5° to 30° in elevation, and care must be taken that this limited field of view does not impose a bias on $X_{\max}$ measurements.

To avoid such a bias, an event is included only if its geometry is such that $X_{\max}$ could be seen and measured at any slant depth between 500 and $1000\,\mathrm{g\,cm^{-2}}$. This is a "conservative" cut removing more events than necessary. The optimum slant depths for this cut are energy dependent and they are selected according to the observed $X_{\max}$ distribution at the corresponding energy . We have experimented with the conservative and the optimum choices and obtained consistent results. To maximize the statistics, we have used the optimum choice for the slant depth limits.

In addition to the cuts described above, we have applied preselection criteria. We excluded data taken during bad calibration periods, or when no information on the atmospheric aerosol content was available, or when the fraction of clouds above the array, as estimated from the LIDARs, was above 25%.

The systematic uncertainty on $\langle X_{\max} \rangle$ becomes larger below $10^{18}$ eV, so we will only present results for energies above $10^{18}$ eV. Measurements of $\langle X_{\max} \rangle$ and its fluctuations below $10^{18}$ eV will soon be obtained using HEAT [12], the new set of fluorescence telescopes installed at the Auger Observatory which view an elevation range from 30° to 60°.

## III. Results

The results of this analysis will be reported at the conference. We will present our measurements of the mean and RMS of the $X_{\max}$ distribution as a function of energy with data collected until March 2009.

# Study of the nuclear mass composition of UHECR with the surface detectors of the Pierre Auger Observatory

**Hernan Wahlberg *, for the Pierre Auger Collaboration†**

* *Physics Department, Universidad Nacional de La Plata C.C. 67-1900 La Plata, Argentina*
† *Pierre Auger Observatory, Av. San Martin Norte 304, (5613) Malargüe, Prov. De Mendoza, Argentina*

***Abstract.*** **We investigate observables that can be measured with the water-Cherenkov detectors of the Pierre Auger Observatory. In particular we explore the use of the risetime of the signals in the detectors and the azimuthal features of the time distributions. A correlation of these observables with the position of shower maximum ($X_{\max}$), as measured with the fluorescence telescopes, is obtained.**

## I. Introduction

The Surface Detector Array (SD) of the southern site of the Pierre Auger Observatory [1] consists of 1660 detectors equally spaced on a triangular grid (1.5 km) over an area of approximately 3000 km$^2$. Each SD detector is a water-Cherenkov detector, with electronics that digitises the signals at 40 MHz sampling rate. The Fluorescence Detector (FD) consists of 4 sites with 6 telescopes each located at the border of the SD array overlooking it. The SD records the shower front by sampling the secondary particles at ground level with a duty close to 100%. The FD measures the fluorescence light emitted as the shower develops through the atmosphere. As it can only operate on clear, moonless nights, its duty cycle is about 13%. FD events provide a direct measurement of $X_{\max}$ ([2] and [3]) that, at present, is the main parameter used to infer mass composition. The bulk of events collected at the Observatory have information only from the surface array and therefore observables from SD, as the ones presented in this paper, are important for composition analysis of Ultra High Energy Cosmic Rays (UHECR).

## II. The risetime of the signal

The time profile of particles reaching ground is sensitive to cascade development as the higher the production height the narrower is the time pulse [4]. The first portion of the signal is dominated by the muon ($\mu$) component which arrives earlier and over a period of time shorter than the electromagnetic particles ($em$).

The risetime ($t_{1/2}$) defined as the time to go from 10% to 50% of the total integrated signal in each station, was shown to be effective for mass discrimination. This is because it is sensitive to the $\mu$ to $em$ ratio, a parameter that varies with the primary mass composition, and is highly correlated with the shower development and the depth of its maximum [5].

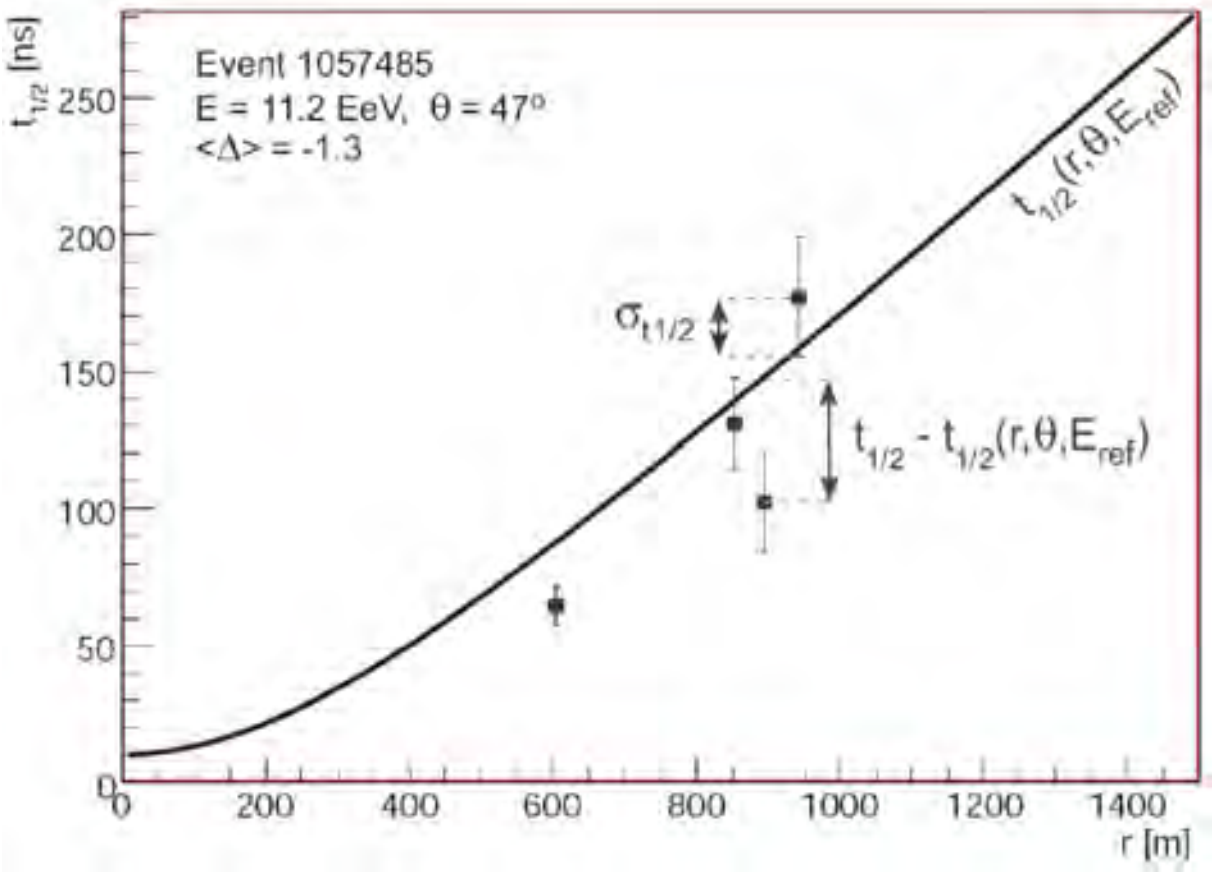

Fig. 1. Risetime vs distance to the core. The curve is the benchmark risetime and the data points represent the measurements of risetime of each detector with uncertainties for this particular event.

A method to obtain the $X_{\max}$ value based on SD observables has been developed. This method consists of obtaining the average value of the risetime as a function of the core distance ($r$) and the zenith angle ($\theta$) for a given reference energy ($10^{19}$ eV), the so-called benchmark. Then, for each selected detector in a given event, the deviation of the measured risetime from the benchmark function is calculated in units of measurement uncertainty and averaged for all detectors in the event as shown in equation 1 and Figure 1, enabling a new observable, $< \Delta_i >$ to be introduced.

$$< \Delta_i >= \frac{1}{N}\sum_{i=1}^{N}\frac{t^i_{1/2} - t_{1/2}(r,\theta,E_{ref})}{\sigma^i_{1/2}(\theta,r,S)}, \qquad (1)$$

where $\sigma^i_{1/2}(\theta,r,S)$ stands for the uncertainty parameterised as function of zenith angle, distance to the core and signal ($S$) of each detector. The $< \Delta_i >$ are expected to be larger for showers developing deeper in the atmosphere than the reference risetime. Figure 2 reflects this fact as the $< \Delta_i >$ is found to increase with energy which is expected as the showers become more penetrating. This parameter has the advantage that can be calculated without any functional adjustment on an event-by-event basis and also it can be determined in events with only one detector satisfying the selection criteria. It is clear from Figure 2 that the rate of change of $< \Delta_i >$ with energy is greater between $3.10^{18}$ and $8.10^{18}$ eV than it is above. Using hybrid events it can be shown that $< \Delta_i >$ is linearly proportional to $X_{\max}$ (Figure 3), confirming the conclusion reached in [5] from simulations. To improve the accuracy of the correlation, signals for each individual detector are deconvolved using single particle response of the electronics. At present the uncertainties are quite large and calibration of the depth parameter based on risetime is on-going. The results shown at the end in Figure 7 are thus to be regarded as preliminary.

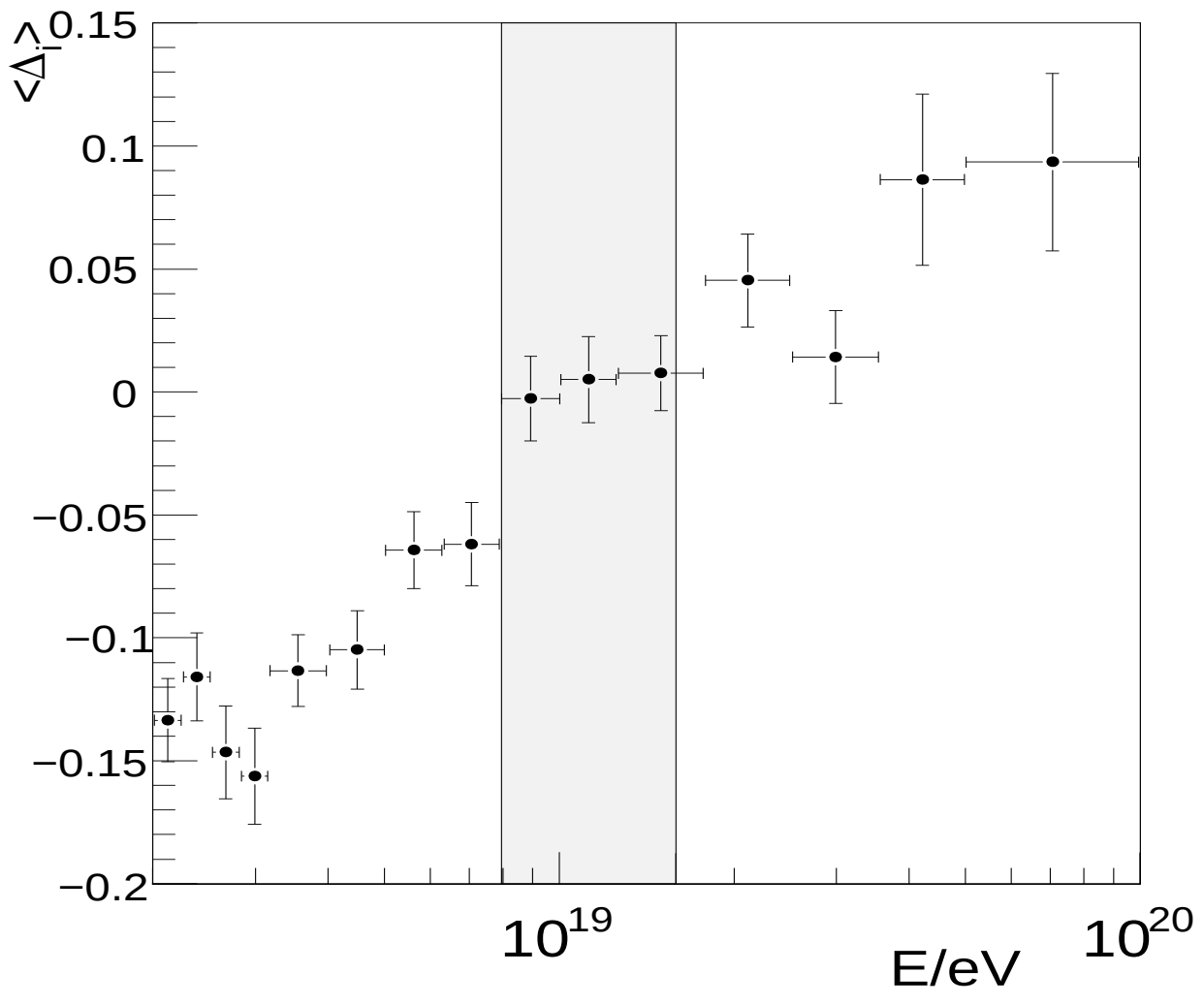

Fig. 2. The average $<\Delta_i>$ as a function of energy for SD events. The dashed lines enclose the region defined for the benchmark function.

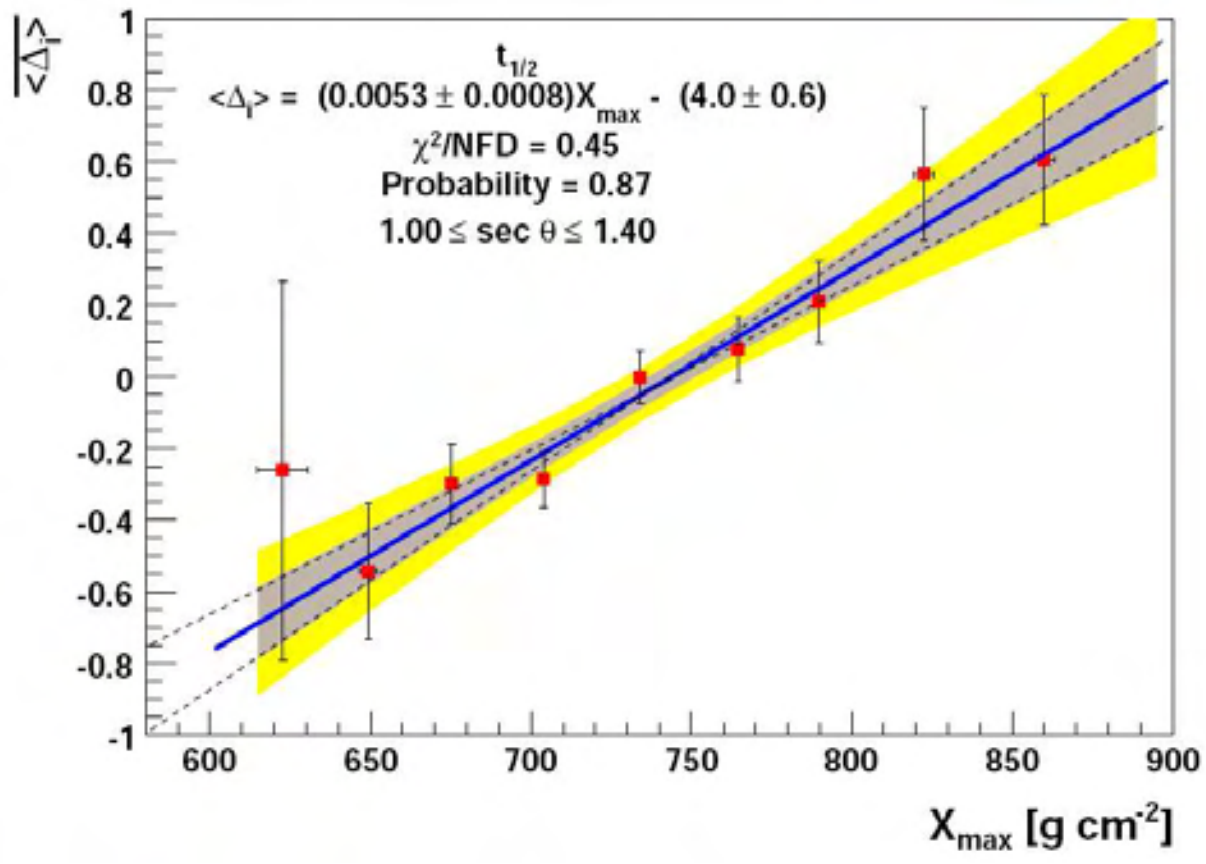

Fig. 3. The average $<\Delta_i>$ as a function of $X_{\max}$ for selected hybrid events. A correlation is found which is parameterised with a linear fit. The shaded areas show the estimated uncertainty (one and two $\sigma$), obtained by fluctuating each point randomly within the measured error bar and repeating the fitting procedure.

## III. Asymmetry in the shower development

The azimuthal asymmetry of time distributions from SD detector signals of non-vertical showers carries valuable information related to the chemical composition of cosmic rays ([6] and [7]).

The risetime asymmetry can be measured by selecting events in bins of energy and $\sec\theta$. Then, for these events the average risetime[1] of those detectors passing quality cuts is determined. After that, for each $(E, \sec\theta)$ bin, a fit of $< t_{1/2}/r >$ to a linear cosine function of $\zeta$ (azimuthal angle in the shower plane) provides the asymmetry factor $b/a$ from:

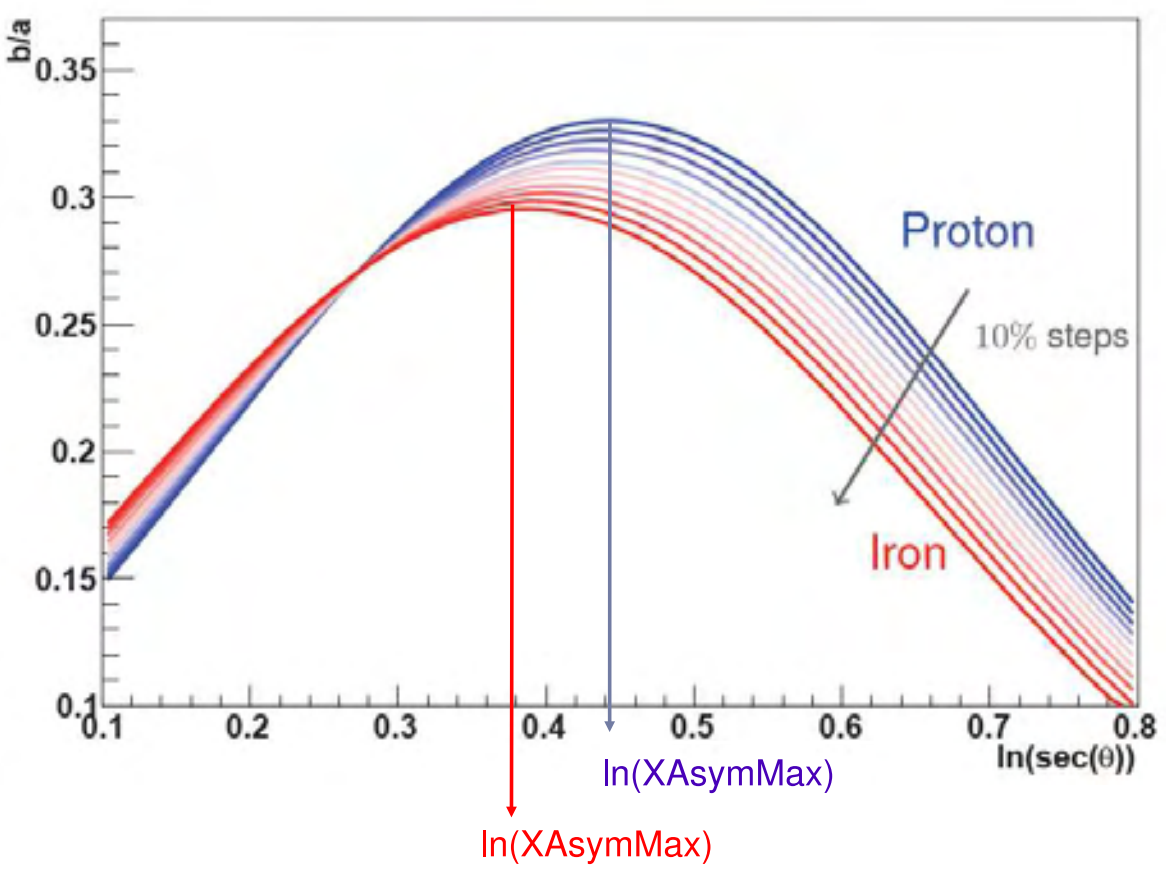

Fig. 4. Asymmetry development for the different samples with mixed composition, going from pure proton to pure iron in steps of 10%. The positions of the maxima for the different primaries are marked.

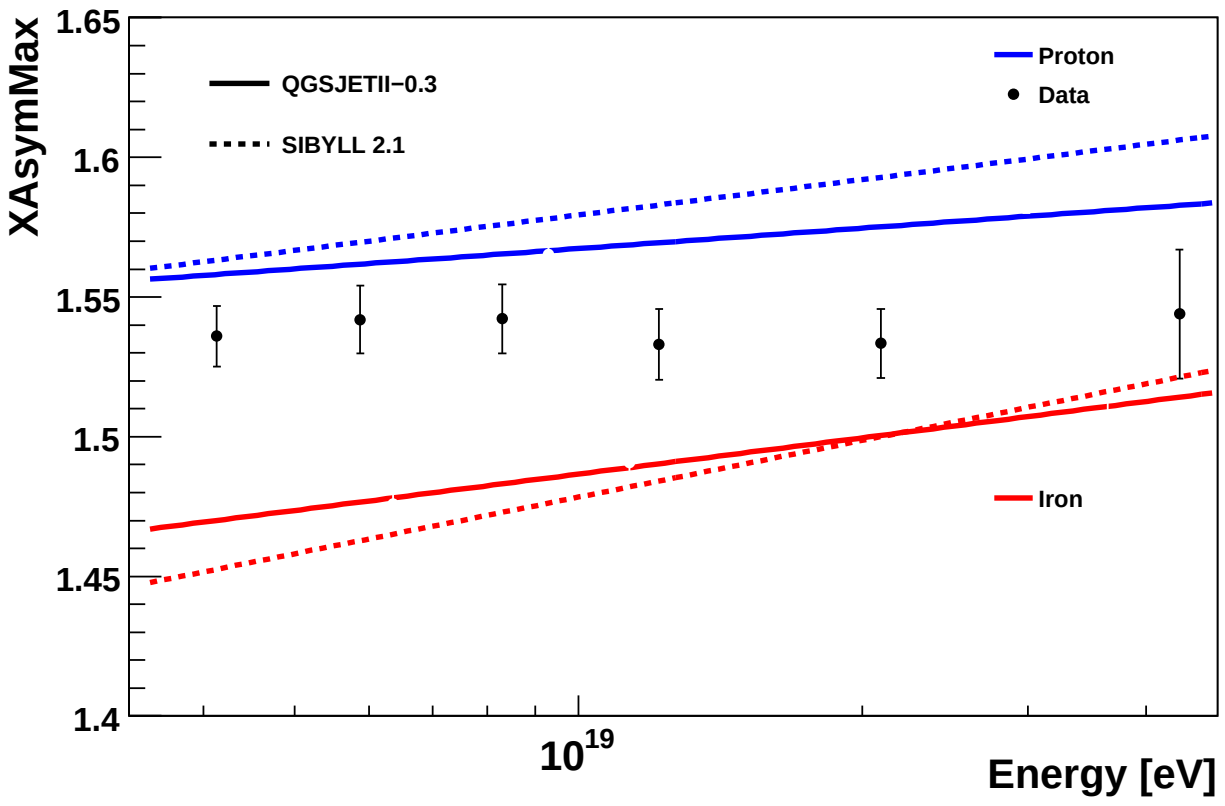

Fig. 5. Position of maximum asymmetry vs. primary energy for different models and primaries. Lines correspond to fitted distributions of MC samples for proton (blue) and iron (red) primaries.

$$< t_{1/2}/r >= a + b\cos\zeta \quad (2)$$

The evolution of $b/a$ with zenith angle is an indicator of the shower development and is different for different primaries as shown in Figure 4. It is worth remarking here, that this method is not based on event-by-event values but is determined by the zenith angle evolution of events grouped in certain energy bins, where a unique value of the asymmetry parameter is obtained for all of them.

In Figure 5 the values of the position $(\sec\theta)$ at which the asymmetry longitudinal development reaches its maximum (XAsymMax) are plotted vs. primary energy for data collected by the Pierre Auger Observatory. Predictions for SIBYLL2.1 and QGSJETII03 hadronic models are included.

[1]As the $t_{1/2}$ increases with the core-distance, $t_{1/2}/r$ is more suitable for asymmetry studies.

XAsymMax, is a robust parameter, only slightly dependent on the number of muons at ground. Hence, a possible change in the muon number predictions from models [9] is not expected to introduce significant changes in the mass composition analysis.

The corresponding linear fits of both primary types are clearly separated, thus allowing discrimination of heavy and light primaries.

As for the parameter $< \Delta_i >$, a calibration with $< X_{\max} >$ can be obtained as shown in Figure 6. In addition, the consistency between MC and data and the universality of these correlations were studied. All the calibration curves are in good agreement within the current statistical uncertainties [8].

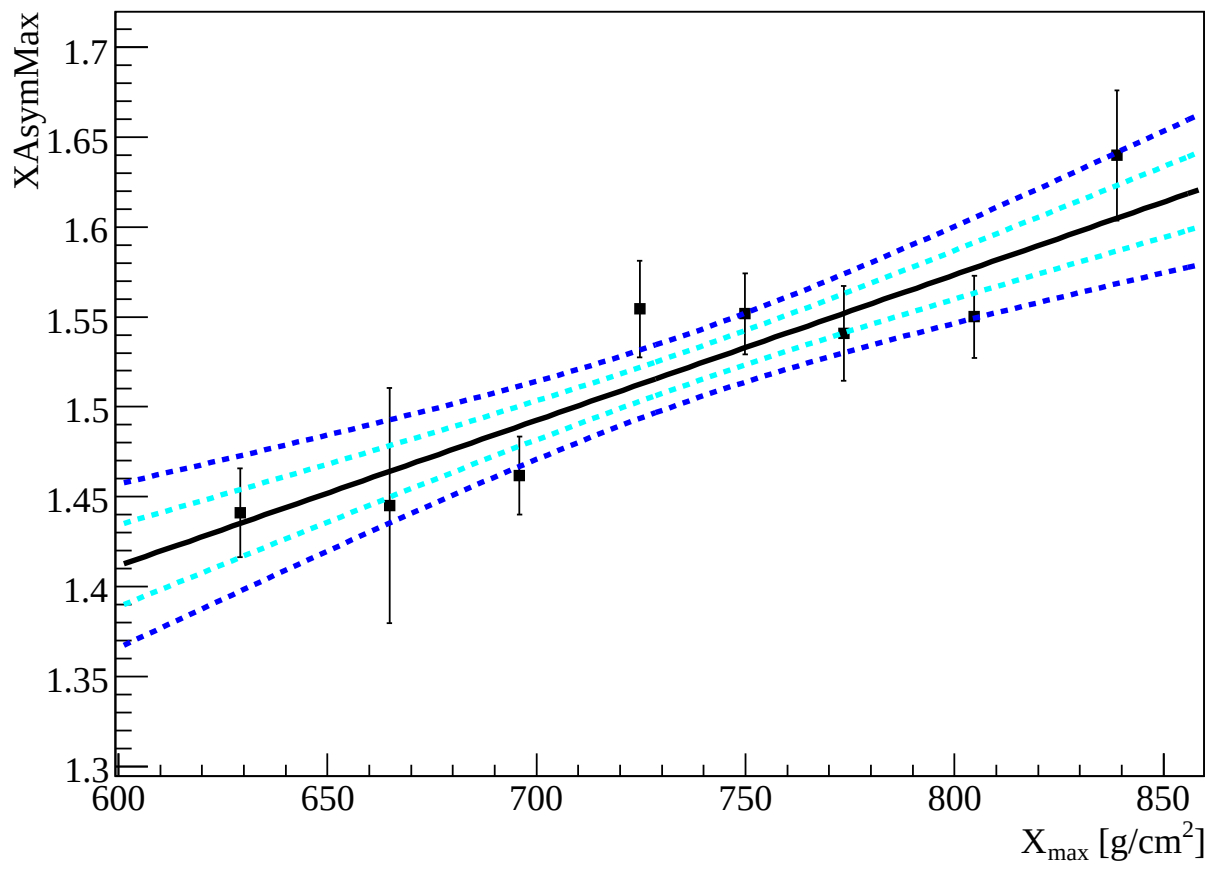

Fig. 6. Calibration curve for data (solid line). Maximum variations (one and two $\sigma$) of the calibration curve when the uncertainties on both fitted parameters are propagated are shown as dashed lines. XAsymMax = a + b $X_{\max}$ with a = (0.84 ± 0.18) and b = (9±2)$10^{-4}$cm$^2$/g.

## IV. Results and Discussion

With present statistics, the systematic uncertainty in $X_{\max}$ obtained due to the parameterisation of the calibration curves are found to be approximately 10 and 16 g cm$^{-2}$ for the risetime and asymmetry methods respectively. The systematic uncertainties are estimated evaluating the half of the variation of $X_{max}$ within the region defined by one $\sigma$ limit curves as shown in Figures 3 and 6.

Figure 7 shows the elongation rate results obtained with both the $< \Delta_i >$ and XAsymMax parameters compared with MC predictions and FD measurements [2]. The results are shown above $3.10^{18}$ eV, the energy at which the surface detector trigger becomes full efficient for both proton and iron primaries.

Both Figures 7 and 5 (obtained only from SD data) suggest that the mean mass increases with energy.

In addition to the parameters presented above, there are additional approaches to mass composition from SD signals currently under study by the Pierre Auger Collaboration. One of them consists in defining the risetime at 1000 m from the core for each event. The other one use the first portion of the signal, meaning the time to reach from 10% to 30% ($t_{10-30}$) of the total integrated signal in each station. The approach based on the risetime at 1000 m defines a $\Delta(1000)$ but with different benchmarks for different energies. The $t_{10-30}$ is more muon dominated and then may show smaller fluctuations and less sensitivity to asymmetry corrections are expected. Both parameters reach a compatible precision but without the need of any deconvolution of the signal allowing less stringent selection of the surface detector units.

In summary, we have shown the sensitivity of the SD array for determining mass composition with two different approaches. One from pure SD measurements as shown in Figure 5. For the other one the SD array is used to determine $X_{\max}$, as shown in Figure 7, from a calibration based on events reconstructed by both SD and FD detectors. Both results are compatible with composition trends indicated from the direct measurements of $X_{\max}$ from the FD detectors.

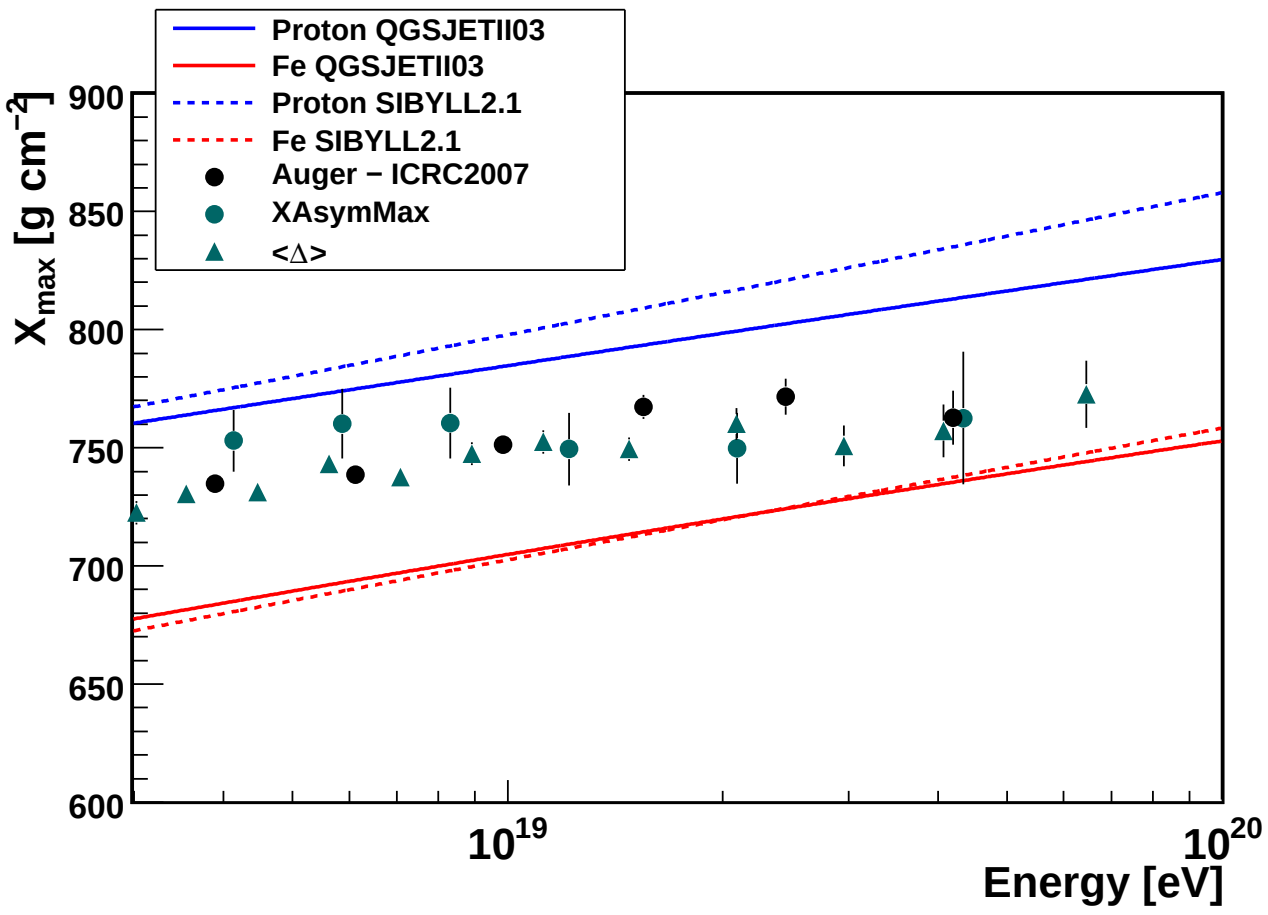

Fig. 7. $X_{\max}$ vs. Energy for both parameters. Predictions for a pure iron and pure proton composition according to different models as well as results from direct measurement of $X_{\max}$ using the FD [2] are shown for comparison. Uncertainties are only statistical.

# Comparison of data from the Pierre Auger Observatory with predictions from air shower simulations: testing models of hadronic interactions

**Antonella Castellina*, for the Pierre Auger Collaboration†**

*Istituto di Fisica dello Spazio Interplanetario (INAF), Università and Sezione INFN, Torino, Italy
†Observatorio Pierre Auger, Av.San Martín Norte 304, (5613) Malargüe, Mendoza, Argentina

***Abstract*. The Pierre Auger Observatory is a hybrid instrument that records the longitudinal, lateral and temporal particle distributions of very high-energy air showers and is sensitive to their electromagnetic and muonic components. Such observables depend on energy and on the type of primary particle that initiates the shower and are sensitive to the hadronic interaction properties. Independent analyses of the combined distributions and direct tests of the predictions of hadronic interaction models are performed at $\simeq 10^{19}$ eV, which corresponds to $\sqrt{s} \simeq 140$ TeV for proton primaries.**

## I. Introduction

The Pierre Auger Observatory is uniquely configured for the investigation of extensive air showers (EAS): with the Fluorescence Detector (FD), we record the longitudinal shower development and measure the shower maximum and the primary energy, while the muonic and electromagnetic components can be measured at ground by the Surface Detector (SD). This information can be used to directly test the predictions of air shower simulations, which due to the indirect nature of EAS measurements are often needed for the interpretation of EAS data. In general, good overall agreement between simulations and measurements is obtained with modern interaction models, but the limits in the modeling of the very high energy hadronic interactions have long been recognized as the largest source of uncertainty [1]. On the other hand, cosmic rays can offer unique information on these interactions in an energy and phase space region not accessible to man-made accelerators.

In this work, we will test the predictions of hadronic interaction models by (a) measuring the muon content of the showers, both by a global method exploiting the shower universality features and by analyses of the temporal particle distributions in the SD and (b) performing direct tests on the simulation of individual hybrid events detected by the Pierre Auger Observatory. The results presented here are based on the data collected with the Pierre Auger Observatory from January 2004 to December 2008. They extend the analysis of [2] to a larger data set and additional, independent analysis methods.

## II. $N_\mu$ measurement using air shower universality

The universality of high-energy showers allows us to describe the surface detector signal at a lateral distance of 1000 m from the core as function of the primary energy $E$, stage of shower evolution $DX \equiv X - X_{\rm max}$, and overall normalization of the muon content [3]. This universality holds to ∼10% for QGSJET II [5] and SIBYLL 2.1 [6] as high-energy interaction models. Denoting the electromagnetic signal by $S_{\rm EM}$ and the muon signal by $S_\mu$, whose evolution with shower age is universal, one can write

$$S_{\rm MC}(E,\theta,< X_{\rm max} >) = S_{\rm EM}(E,\theta,DX) + N_\mu^{\rm rel} S_\mu^{\rm QGSII,p}(10^{19}\,{\rm eV},\theta,DX) \quad (1)$$

where $N_\mu^{\rm rel}$ is defined as the number of muons relative to that of QGSJET II proton showers at $10^{19}$ eV and $S_\mu^{\rm QGSII,p}$ is the muon signal predicted by QGSJET II for proton primaries. Since $\langle X_{\rm max}\rangle$ is known from FD measurements, the only unknown in Eq.(1) is $N_\mu^{\rm rel}$, which can be measured at a reference energy $E_0 =$ 10 EeV using the isotropy of the cosmic ray flux and the angular dependence of $S_{\rm MC}(E_0,\theta)$ through $N_\mu^{\rm rel}$ [2], [4]. Analyzing the full data set, the muon number relative to proton-QGSJET II is
$N_\mu^{\rm rel}(10\,{\rm EeV}) = 1.53^{+0.09}_{-0.07}$ (stat.)$^{+0.21}_{-0.11}$ (syst.).
The systematic error includes the remaining primary particle-dependence of the electromagnetic signal as well as the effect of shower-to-shower fluctuations and the uncertainty on $\langle X_{\rm max}\rangle$. Knowing $N_\mu^{\rm rel}(10\,{\rm EeV})$ and the measured mean depth of shower maximum, the signal size at $\theta = 38°$ can be calculated
$S_{38}(10\,{\rm EeV}) = 38.9^{+1.4}_{-1.2}$ (stat.)$^{+1.6}_{-1.8}$ (syst.) VEM,
which corresponds to assigning showers a 26% higher energy than that of the current FD calibration [7].

## III. $N_\mu$ from the FADC traces

The separation of the muonic and electromagnetic components of the SD signal relies on the FADC traces recorded by each of the 3 PMTs of the SD detectors. Each trace is sampled at a frequency of 40 MHz [8]. As a typical muon from a UHECR shower deposits much more energy ($\approx 240\,{\rm MeV}$) in a water tank than an electron or photon ($\lesssim 10\,\mathrm{MeV}$), spikes are produced over the smoother electromagnetic background in the FADC time traces, see Fig. 1. Thus, muons manifest themselves as sudden variations in the signal. High-energy photons in a shower can produce a sudden increase of the electromagnetic signal similar to that of a muon: their contribution is estimated to be $\lesssim 10\%$.

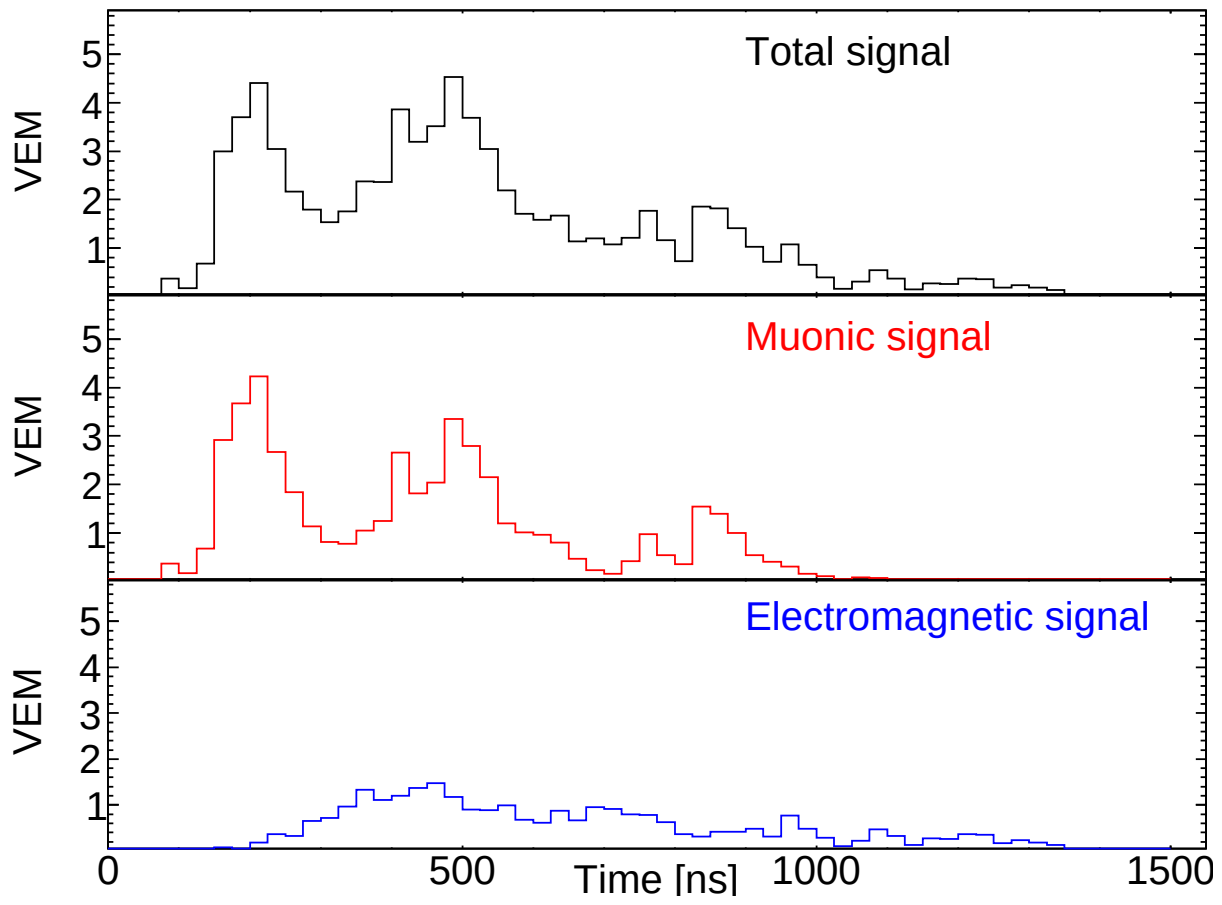

Fig. 1: Example of a simulated FADC trace in one station of the surface detector. From top to bottom: total trace, muonic and EM components. Shown is the tank signal from a proton shower of $\theta = 45°$ and $E = 10^{19}\,(eV)$ at $1000\,\mathrm{m}$ from the core.

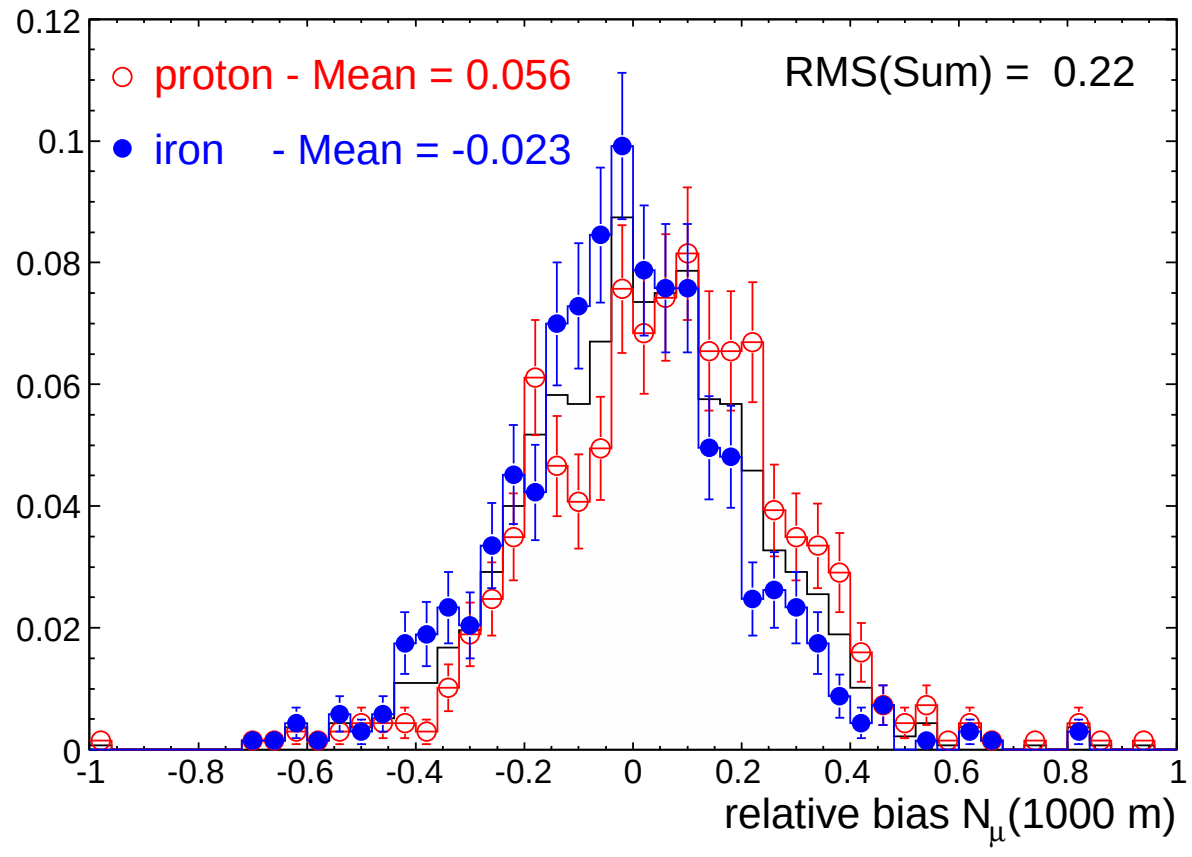

Fig. 2: Dependence of the relative difference between the simulated and the estimated number of muons on the primary particle. The results are presented for 10 EeV energy showers simulated with QGSJET II and zenith angles up to 50°.

### *A. The jump method*

To extract muon spikes, we define the FADC jump $v$ as the difference of FADC values of two consecutive time bins [9]. The main idea is that of evaluating the sum of the jumps larger than a threshold $v_{\mathrm{thr}}$ which is determined by finding the best compromise between muon selection efficiency and electromagnetic contamination. The raw jump integral

$$\mathcal{J}(v_{\mathrm{thr}}) = \int_{v_{\mathrm{thr}}}^{\infty} \left(\frac{\mathrm{d}N}{\mathrm{d}v}\right) v\,\mathrm{d}v = \sum_{\substack{v>v_{\mathrm{thr}} \\ t_i,\mathrm{ADC\ bin}}}^{\infty} v(t_i) \qquad (2)$$

is then corrected to calibrate our estimator in terms of number of muons by a factor $\eta(E, \theta, r)$ which depends on the primary energy $E$, the zenith angle $\theta$, and the distance $r$ of the detector from the shower core.
Monte Carlo simulations based on CORSIKA [10] were used to derive the dependences of the correction factor on the shower parameters and to estimate the possible bias for ultra-high energies and for distances close to the core. The number of muons at 1000 m is determined with a resolution close to 20% and systematic biases below 7%. The relative difference between the simulated and the estimated number of muons is shown in Fig. 2 for different primary particles.

### *B. The smoothing method*

The electromagnetic (EM) contribution to the signal in the surface detectors can be estimated by a smoothing method. The total trace recorded by the FADCs of the 3 PMTs in each station is averaged over a preset number of consecutive time bins $N_{\mathrm{bin}}$. Any positive difference between the original trace and the smoothed signal is assigned to the muon component and subtracted from the signal; then the whole procedure is applied again for a number of iterations $N_{\mathrm{iter}}$. Using Monte Carlo simulations, the best parameters $[N_{\mathrm{bin}}, N_{\mathrm{iter}}]$ are determined as those minimizing the bias in the evaluation of the EM component for both proton and iron primaries, and for the largest possible angular range.
Based on simulations, a correction factor $\xi(E, \theta, r)$ is determined that depends on the energy of the primary particle, its zenith angle and the distance to the shower core. For $E > 3$ EeV (full efficiency of the SD) and distances around 1000 m from the core, the EM contribution to the signals is evaluated with a resolution of 23% and a systematic uncertainty below 8%, irrespectively of the primary energy and composition. The relative difference between the estimated and expected EM signals is shown in Fig. 3.

The relative difference $\Delta(S_{\mathrm{EM}}/E)$ between the EM signal reconstructed from data and the QGSJET II prediction assuming all primaries are protons (red empty symbols) or iron (blue filled circles) is shown in Fig. 4. Due to an almost linear energy scaling of the EM shower signal, this discrepancy could be removed by assigning showers a 29% higher energy than from FD calibration. Alternatively, the discrepancy could also be related to an incorrect description of the lateral distribution of EM particles in the simulation.
The muon component in each detector is derived by difference, after having evaluated the EM one, with systematic uncertainties below 8% and a resolution close to 20%.

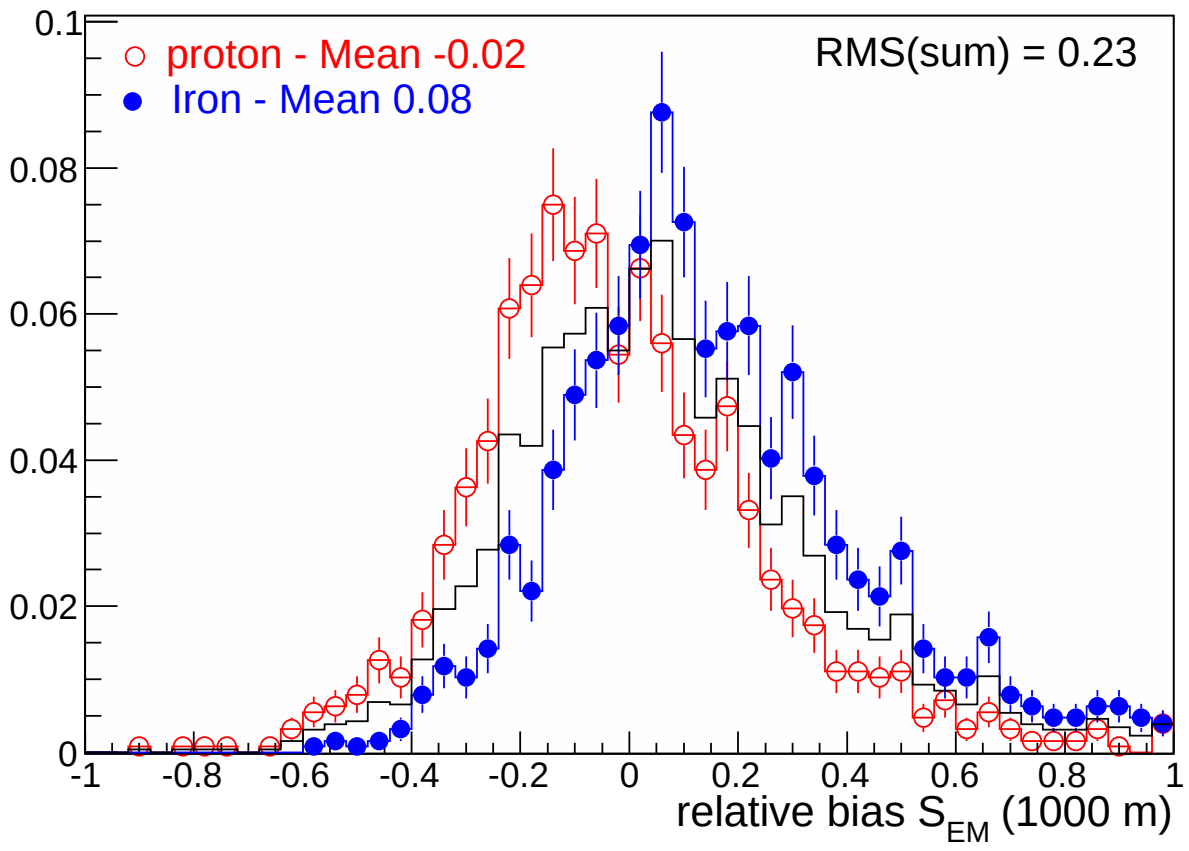

Fig. 3: The dependence of the relative deviation between the simulated and the estimated EM signals on the primary particle. The results are presented for 10 EeV energy showers and zenith angles up to 50°.

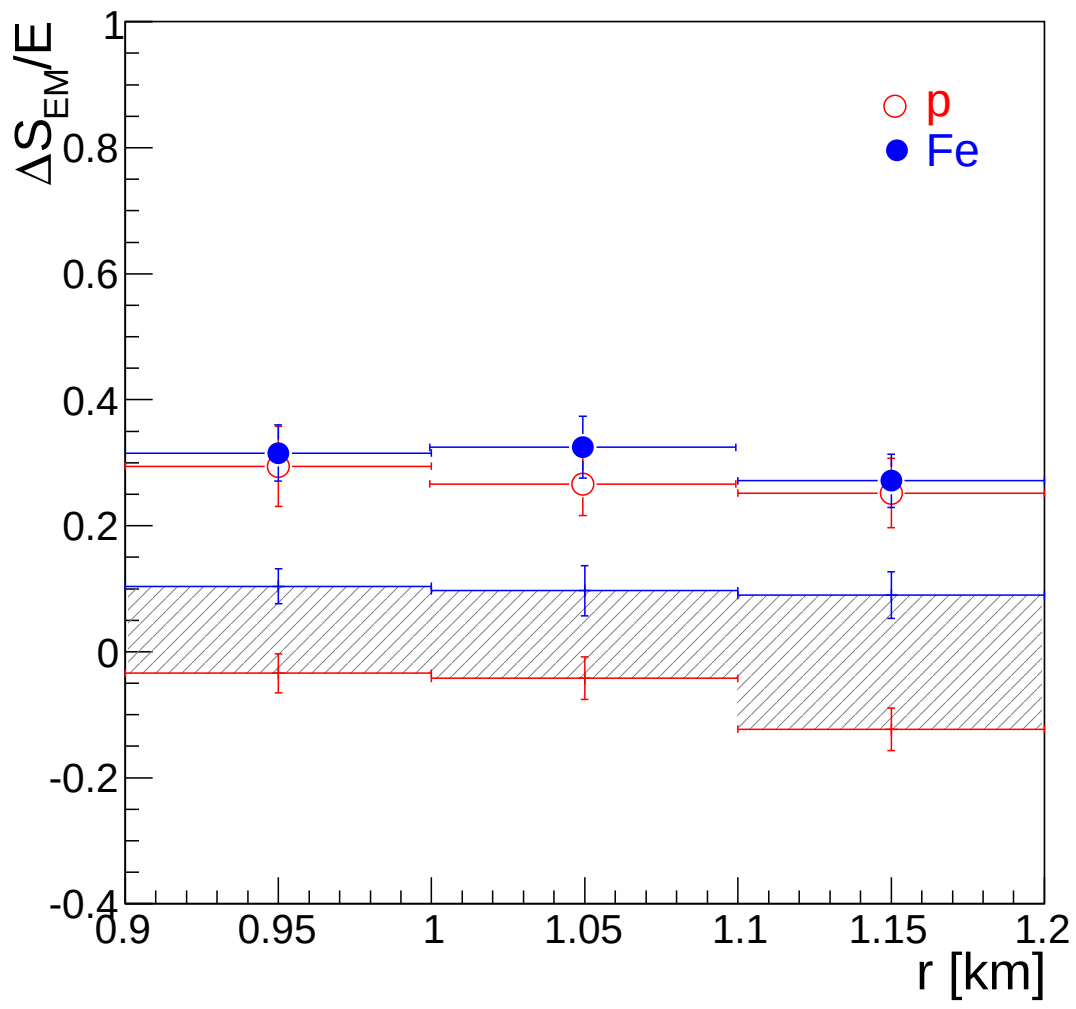

Fig. 4: The relative difference between the EM signals in data and in the simulation (open and filled symbols indicate the use of proton or iron primaries in the simulation, respectively). The systematic uncertainty for $S_{\rm EM}$ (10 EeV and 38° showers) is shown by the shaded band.

## IV. Individual Hybrid Simulation

The FD and SD signals can be compared to the model predictions on an event-by-event basis with a technique based on the simulation of individual high quality hybrid events. The shower simulations are done using SENECA [11] and QGSJET II as high energy hadronic interaction model. The surface detector response has been simulated with GEANT4 and extensively tested [12]. We use hybrid events with $18.8 < \log_{10}(E/\text{eV}) < 19.2$ that satisfy the quality cuts used in the FD-SD energy calibration and $X_{\rm max}$ analyses [7], [13]. Each event is at first simulated 400 times using the geometry and energy given by the hybrid reconstruction of the event. The primary is taken as proton or iron as is most probable based on the measured $X_{\rm max}$ of the event. The three simulated showers with the lowest $\chi^2$ with respect to the FD data are then re-simulated using a lower thinning level to have a high quality simulation of the particles reaching ground. Finally, the actual detector response to each of the simulated events is obtained using [14]. The longitudinal profiles and the lateral distribution functions variation among the three simulations is $\approx 5$ and 15%, respectively . The measured longitudinal profile together with that of the best-matching simulated event is shown in Fig. 5 (top panel) for one representative event; in the bottom panel, the measured tank signals are compared to those of the simulated event.

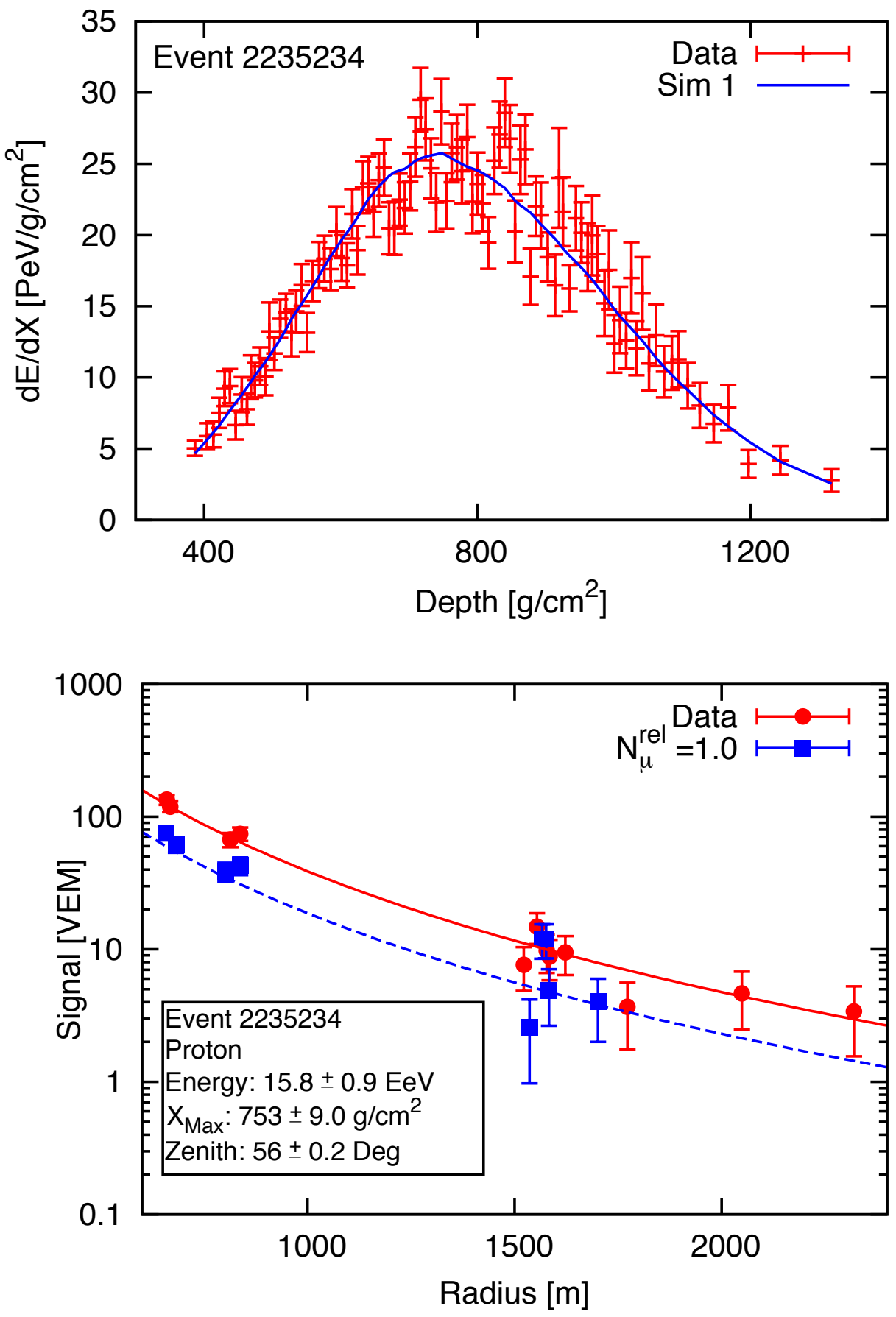

Fig. 5: The observed longitudinal (top panel) and lateral (bottom panel) profiles for one of the hybrid events. The best-matching simulation is shown by the full (top) and dashed (bottom) line (without rescaling of the muon number relative to the model prediction).

An overall rescaling of the surface detector signals results in a residual discrepancy which increases approximately linearly with sec$\theta$ of the events; a possible interpretation of this deficit of signal is a lack of muons in the simulations. The preferred energy and muon shift within the Golden Hybrid method can be found determining for each event the reconstructed $S(1000)$, as a function of the EM and muonic renormalizations, by performing the detector simulations and event reconstruction with individual particle weights adjusted according to the rescaled values. The best rescaling is taken to be that which minimizes the $\chi^2$ between simulated and observed $S(1000)$'s for the ensemble of events. The "one-$\sigma$" contour is found by propagating the statistical uncertainties from the best fit as well as the systematic uncertainties. As can be seen in Fig. 6, there is a strong correlation between the two parameters and the $\chi^2$ minimum is quite broad.

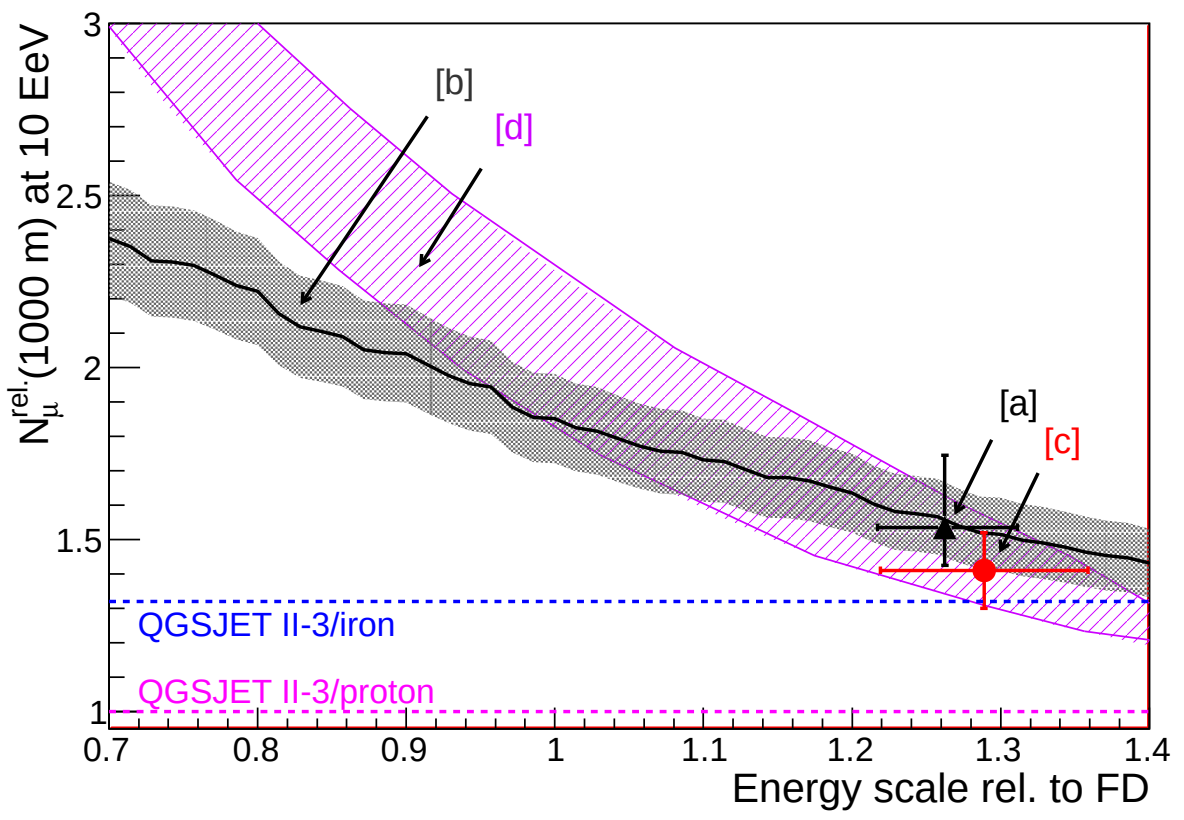

Fig. 6: Number of muons at 1000 m relative to QGSJET-II/proton vs. the energy scale from [a] the universality method (triangle); [b] the jump method (filled area); [c] the smoothing method (circle); [d] the golden hybrid analysis (dashed area). The events have been selected for $\log_{10}(E/\mathrm{eV}) = 19.0 \pm 0.02$ and $\theta \leq 50°$. According to the tested model, Iron primaries give a number of muons 1.32 times bigger than that from protons (horizontal lines in the figure).

## V. Results

The derived number of muons relative to that predicted by QGSJET-II for proton primaries and the energy scale with respect to the fluorescence detector are shown in Fig. 6 for 10 EeV primaries with zenith angle below $50°$. The results of all four analysis methods are compatible with each other. The analysis based on shower universality yields a measure of the energy scale which is almost independent of the fluorescence detector calibration, fixing it to $E' = 1.26^{+0.05}_{-0.04}(\mathrm{syst.}) \times E_{FD}$; the smoothing technique constrains the relative energy scale to a value $E' = (1.29 \pm 0.07\,(\mathrm{syst.})) \times E_{FD}$ from the analysis of the electromagnetic signals alone. The two energy scales agree with each other and are compatible with the currently used FD energy assignment that has a systematic uncertainty of 22% [15]. Adopting the energy scale $E'$, the analyses agree with the conclusion of a muon signal 1.3÷1.7 times higher than that predicted by QGSJET-II for protons. With this energy scale, the results indicate a muon deficit in simulated showers, being only marginally compatible with the prediction of QGSJET-II for primary iron ($N_\mu^{\mathrm{rel}}$=1.32) within the systematic uncertainties of the different methods used to derive the muon contribution. The observed mean and distribution of the depth of maximum of the showers, however, is clearly at variance with the predictions of QGSJET or SIBYLL for a pure iron composition [16].

The results presented here are obtained at a lateral distance of $1000\,\mathrm{m}$. The rescaling factor found for the muon density does not necessarily apply to the total number of muons in a shower as it is not known how well the models reproduce the lateral distribution of muons. Moreover, the energy scale found in this work is based on the assumption of a correct reproduction of the lateral distribution of EM particles by simulations made with EGS4 [17] in combination with the hadronic models QGSJET and SIBYLL.

Recent work on hadronic interactions [18] has shown that an increase of the predicted muon number of EAS can be obtained if the description of baryon-pair production in hadronic interactions is modified.

# A Monte Carlo exploration of methods to determine the UHECR composition with the Pierre Auger Observatory

**Domenico D'Urso*** **for the Pierre Auger Collaboration**†

*Dipartimento di Fisica dell'Università di Napoli and Sezione INFN, Via Cintia 2, 80123 Napoli, Italy
†Pierre Auger Observatory, Av. San Martin Norte 304, (5613) Malargüe, Prov. De Mendoza, Argentina

***Abstract*. Measuring the mass composition of ultra high energy comsic rays is crucial for understanding their origin. In this paper, we present three statistical methods for determining the mass composition. The methods compare observables measured with the Pierre Auger Observatory with corresponding Monte Carlo predictions for different mass groups obtained using different hadronic interaction models. The techniques make use of the mean and fluctuations of $X_{max}$, the log-likelihood fit of the $X_{max}$ distributions and the multi-topological analysis of a selection of parameters describing the shower profile. We show their sensitivity to the input composition of simulated samples of known mixing and their ability to reproduce mass sensitive observables, like the average shower maximum as a function of the energy, measured at the Pierre Auger Observatory.**

## I. INTRODUCTION

The understanding of the nature of the ultra high energy cosmic rays (UHECR) is a crucial point towards the determination of their origin, acceleration and propagation mechanisms. The evolution of the energy spectrum and any explanation of its features strongly depend on the cosmic chemical composition since the galactic confinement, the attenuation length of various energy loss mechanisms and the energy achievable at the sources depend on the primary particle type. Above $10^{19}$ eV all observed cosmic particles are presumed to have extragalactic origin, because there are no galactic sources able to produce particles up to such energies and they cannot be confined in our galaxy long enough to be accelerated. The energy at which the transition from galactic to extragalactic cosmic rays occurs is still unknown and only a detailed knowledge of the composition spectrum will allow to discriminate among different astrophysical models [1].

The *hybrid* design of the Auger Observatory, the integration of a surface detector array (SD) and a fluorescence detector (FD), exploits stability of experimental operation, a 100% duty cycle, and a simple determination of the effective aperture for the SD, calorimetric shower detection, direct observation of shower longitudinal profile and shower maximum for the FD. This hybrid design allows to simultaneously use the most sensitive parameters to the primary mass from both the SD and FD: the slant depth position $X_{max}$ at which the maximum of the shower profile is reached and its fluctuations from the FD [2][3], the signal risetime in the Cherenkov stations, the curvature of the shower front, the muon-to-electromagnetic ratio and the azimuthal signal asymmetry from the SD [4]. In this paper, we present three statistical techniques for determining the mass of primary particles. These methods compare shower observables measured by the FD at the Pierre Auger Observatory with corresponding Monte Carlo predictions, including a full detector simulation.

## II. COMPOSITION ANALYSIS WITH THE MOMENTS OF $X_{max}$ DISTRIBUTION

The first two moments of $X_{max}$ distribution, the mean and its variance, have been used as mass discriminators. We derive the mass composition from the best choice of primary fractions that reproduce experimental data using their expectation values (method of moments, MM). With this two observables, the cosmic ray flux can be modeled as a mixture of three primary masses (a, b and c) and define the two parameters describing the mixture $P_1$ and $P_2$. The relative abundances in terms of $P_1$ and $P_2$ are

$$\begin{aligned} \mathrm{P_a} &= P_1 \\ \mathrm{P_b} &= P_2(1-P_1) \\ \mathrm{P_c} &= (1-P_1)(1-P_2) \end{aligned} \quad (1)$$

The expected mean shower maximum of the mixture is

$$\langle \mathrm{X_{exp}} \rangle = \mathrm{P_a}\langle \mathrm{X_a} \rangle + \mathrm{P_b}\langle \mathrm{X_b} \rangle + \mathrm{P_c}\langle \mathrm{X_c} \rangle \quad (2)$$

where $\langle \mathrm{X_a} \rangle, \langle \mathrm{X_b} \rangle \mathrm{and} \langle \mathrm{X_c} \rangle$ are the mean $X_{max}$ for simulated data sets of species a,b and c. The expected $X_{max}$ fluctuations (i.e., the root mean square of the $X_{max}$ distribution) $\Delta X_{exp}$ for a mixture of three masses can be written in an easier way defining first its value for two masses, b and c, and then considering its mixture with the species a:

$$\langle \mathrm{X_{b-c}} \rangle = \mathrm{P_2}\langle \mathrm{X_b} \rangle + (1-\mathrm{P_2})\langle \mathrm{X_c} \rangle; \quad (3)$$

$$(\Delta \mathrm{X_{b-c}})^2 = \mathrm{P_2}\Delta \mathrm{X_b^2} + (1-\mathrm{P_2})\Delta \mathrm{X_c^2} + \mathrm{P_2}(1-\mathrm{P_2})(\langle \mathrm{X_b} \rangle - \langle \mathrm{X_c} \rangle)^2; \quad (4)$$

$$(\Delta \mathrm{X_{exp}})^2 = \mathrm{P_1}\Delta \mathrm{X_a^2} + (1-\mathrm{P_1})\Delta \mathrm{X_{b-c}^2} + \mathrm{P_1}(1-\mathrm{P_1})(\langle \mathrm{X_a} \rangle - \langle \mathrm{X_{b-c}} \rangle)^2; \quad (5)$$

where $\Delta X_i$, $\langle X_{i-j} \rangle$ and $\Delta X_{i-j}$ are the $X_{max}$ fluctuations for the primary i and the mean and its fluctuations for the mixture i-j. Assuming that the data set is so large that $\langle X_{max} \rangle$ and $\Delta X_{max}$ are statistically independent, in each energy bin we can fit the data could be fitted solving equations 5 and 2 with two unknowns, $P_1$ and $P_2$.

## III. Mass Composition from a logarithmic likelihood fit to $X_{max}$ distribution (LLF)

The method assumes that the observed events $N_{data}$ are a mixture of $N_m$ pure mass samples with unknown fractions $p_j$. The expected number of showers $\nu_i$ with $X_{max}$ into i-th bin is therefore:

$$\nu_i(\mathbf{p}) = N_{data} \sum_{j=0}^{N} p_j \frac{a_{ij}}{N_j^{MC}} \quad , \; i = 1, \ldots, N \qquad (6)$$

where $a_{ij}$ are the number of Monte Carlo events from primary j in bin i, N is the total number of bins in the $X_{max}$ distribution and $N_j^{MC} = \sum_{i=0}^{N} a_{ij}$. The probability $P(n_i)$ to observe $n_i$ events in the i-th bin is given by the product of the Poisson distributions of mean $\nu_i$

$$P(n_i) = \prod_{i=1}^{N} \frac{\nu_i^{n_i}}{n_i!} e^{-\nu_i}$$

The logarithm of $P(n_i)$ gives the log-likelihood function:

$$\log L = \sum_{i=0}^{N} [\, n_i \; \log \, \nu_i - \nu_i - \log \, n_i!] \qquad (7)$$

Maximizing eq. 7 with respect to $p_j$, one finds the primary fractions in the measured data sample.

In Fig. 1, we show the $X_{max}$ distribution for a sample of 70% proton-30% iron (black dots) between $10^{18.2}$ and $10^{18.3}$ eV fitted by the weighted sum of the expected $X_{max}$ distribution for proton (dotted line) and iron (dashed line). The techniques searches for the best choice of primary fractions that optimize the fit.

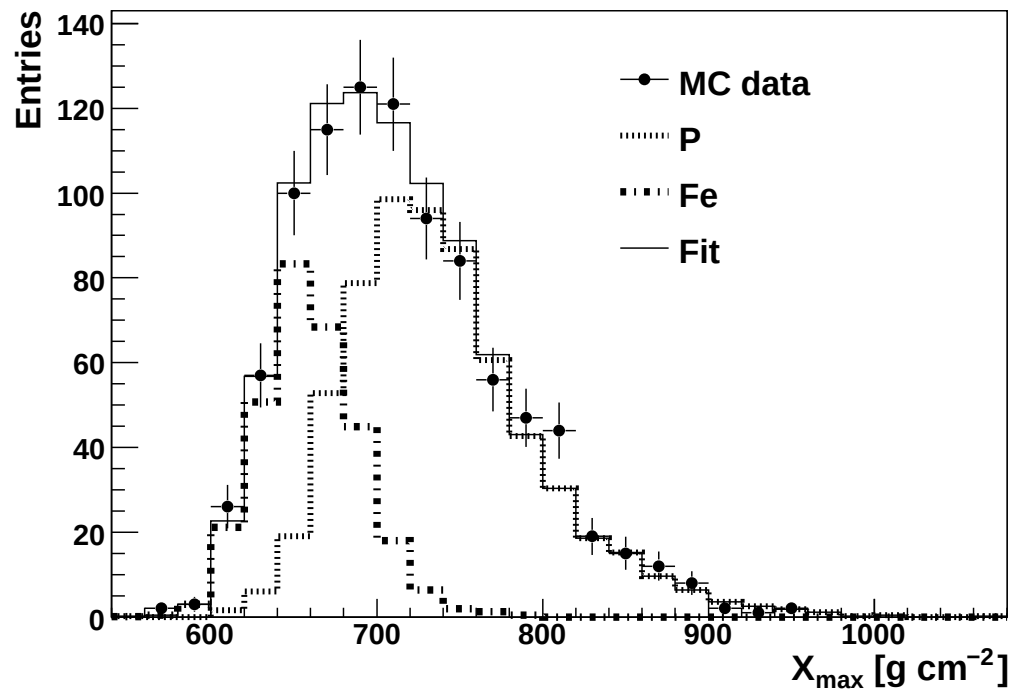

Fig. 1. $X_{max}$ distribution for a sample of 70% proton-30% iron (black dots) between $10^{18.2}$ and $10^{18.3}$ eV fitted by the weighted sum of the expected $X_{max}$ distribution for proton (dotted line) and iron (dashed line) with the best best choice of proton and iron fractions. Distributions are normalized to the number of events in the test sample.

## IV. Multiparametric Analysis for the primary composition

We apply the multiparametric topological analysis (MTA) described in [5] to classify observed showers using the correlations of their characteristics. Starting from a set of observables, it is possible to define a parameter space, which is divided in cells whose dimensions are related to the experimental accuracy. A wide set of simulated cascades produced by different primary nuclei is used to populate the parameter space. In each cell, that in the most general n-dimensional case is defined by $(h_1 \ldots h_n)$, one can define the total number of showers $N_{tot}^{(h_1 \ldots h_n)}$ and the total number of showers induced by the primary i $N_i^{(h_1 \ldots h_n)}$ populating the cell, and then derive the associated frequency:

$$p_i^{(h_1 \ldots h_n)} = N_i^{(h_1 \ldots h_n)} / N_{tot}^{(h_1 \ldots h_n)} \qquad (8)$$

which can be interpreted as the probability for a shower falling into the cell $(h_1 \ldots h_n)$ to be initiated by a nucleus of mass i. Considering a sample of M showers, its fraction of primary j is given by

$$p_j = \sum_{m=1}^{M} p_j^{(h_1 \ldots h_n)_m} / M \qquad (9)$$

with $(h_1 \ldots h_n)_m$ indicating the cell interested in by the m-th event.

A second set of showers is used to compute the mixing probabilities $P_{i \to j}$ that an event of mass i is identified as primary j. The mean $P_{i \to j}$ is obtained by computing $p_j$ for samples of pure primary composition i. Assuming the measured sample as composed by $N_m$ species, the mixing probabilities $P_{i \to j}$ can be used for the reconstruction of the primary mass composition as the coefficients in the system of linear equations:

$$\begin{aligned} N'_1 &= \sum_{i=1}^{N_m} N_i \cdot P_{i \to 1} \\ \vdots &= \vdots \\ N'_{N_m} &= \sum_{i=1}^{N_m} N_i \cdot P_{i \to N_m} \end{aligned} \qquad (10)$$

where $N_i$ are the true values, which are altered to $N'_j$ due to misclassification. The solution of eq. 10 gives the mass composition of the measured sample in terms of $N_m$ primary masses and, dividing by the total number of showers $N_{data}$, their fractions $p_j$.

MTA performances on CONEX [6] showers, fully simulated through the Auger apparatus, has been already presented [5]. Its performances has been also recently tested on a set of observables from the longitudinal profile and the lateral distribution of CORSIKA [7] simulated showers [8].

In this paper, the MTA application to only FD data using 2 parameters is described and the space is defined by $X_{max}$ and $X_0$ of the Gaisser-Hillas function

$$\frac{dE}{dX} = \left(\frac{dE}{dX_{max}}\right)\left(\frac{X - X_0}{X_{max} - X_0}\right)^{\frac{X_{max}-X_0}{\lambda}} e^{\frac{X_{max}-X_0}{\lambda}}$$

where $dE/dX$ and $(dE/dX)_{max}$ are the energy deposit at the depth X and at the shower maximum. In Fig. 2 the parameter space built for ($X_{max}$, $X_0$) between $10^{18.9}$ and $10^{19.1}$ eV. The space is divided in cells with dimensions 20 and 50 g cm$^{-2}$ respectively and is populated with Conex simulated showers induced by proton (dots) and iron (triangles). Clearly the effective parameter is $X_{max}$. Despite that, a 2-parameter case is reported to show how the technique can include N parameters in a natural way. The on-going extention to quantities measured by the SD allows to a larger set of effective parameters and to better discriminate among different primaries.

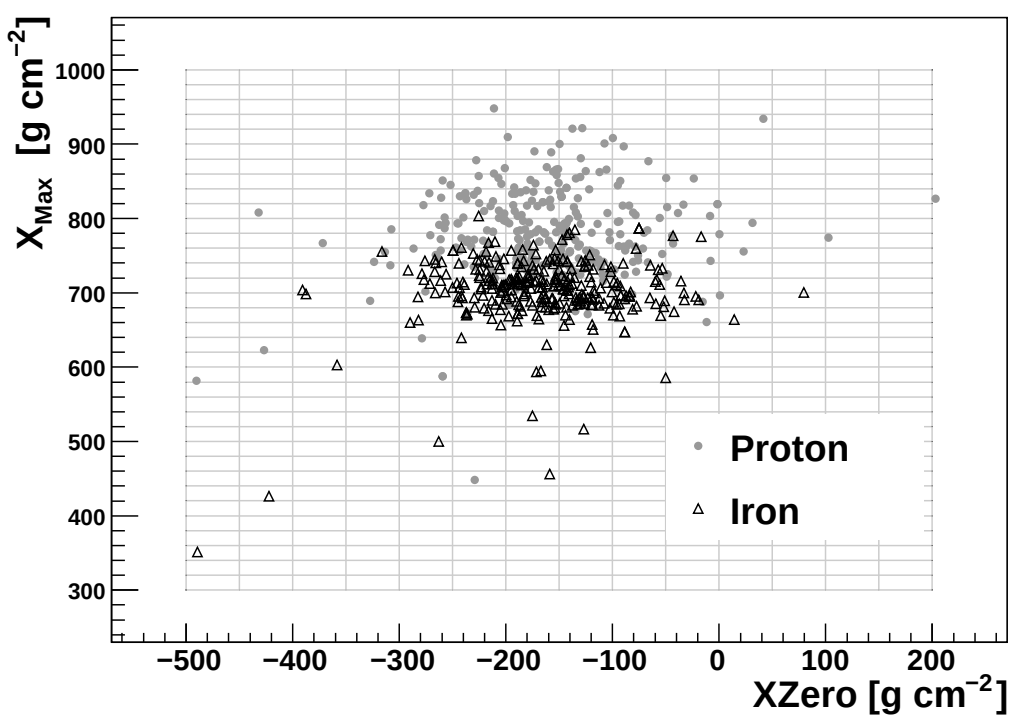

Fig. 2. Parameter space built for ($X_{max}$, $X_0$) between $10^{18.9}$ and $10^{19.1}$ eV. The space is divided in cells with dimensions 20 and 50 g cm$^{-2}$ respectively and is populated with Conex simulated showers induced by proton (dots) and iron (triangles).

## V. Influence of reconstruction efficiency

The reconstruction of event fractions in terms of $N_m$ masses could be altered by different efficiencies with respect to each primary particle (possible trigger, reconstruction and selection effects). The field of view of the Auger FD, located at 1400 m over the sea level (870 g cm$^{-2}$ of vertical depth), covers an elevation angle from 1.5 to 30. If we require to detect the shower maximum to ensure a good $X_{max}$ resolution, we favour light primaries at lowest energies and heavy nuclei at highest energies. To have an unbiased measurement due to the FD field of view limits one should select at each energy only geometry ranges (zenith angle, etc..) at which this effect is negligible (see [2] and [3] for further details).

Such cuts can be avoided, retaining a larger statistics, if the obtained primary fractions are corrected taking into account the reconstructed efficiencies for each primary mass. The reconstruction efficiency can be determined as the ratio between the total number of accepted events over the total number of generated events for the mass j, $\epsilon_j^{tot} = N_j^{acc}/N_j^{gen}$, and for a specific observable in a particular bin i of its distribution the number of expected events should be weighted by $\epsilon_{ji} = N_{ji}^{acc}/N_{ji}^{gen}$. The corrected fractions $p_j^{corr}$ are $(p_j/\epsilon_j^{tot})/(\sum_{j=1}^{N_m} p_j/\epsilon_j^{tot})$.

## VI. Method Performances

The described techniques have been tested on simulated samples of known composition. For different proton-iron mixing, N events have been randomly selected from proton and iron Monte Carlo data and the resulting samples have been analyzed. The whole procedure have been repeated many times. In Fig. 3, the mean value and the root mean square of the distribution of the difference between the reconstructed input fractions and the expected ones are shown for different mixtures of protons and iron CONEX showers, using QGSJETII-03 [9], fully simulated through the Auger detector with the Auger analysis framework [10]. The input abundances are well reproduced by the methods in all cases.

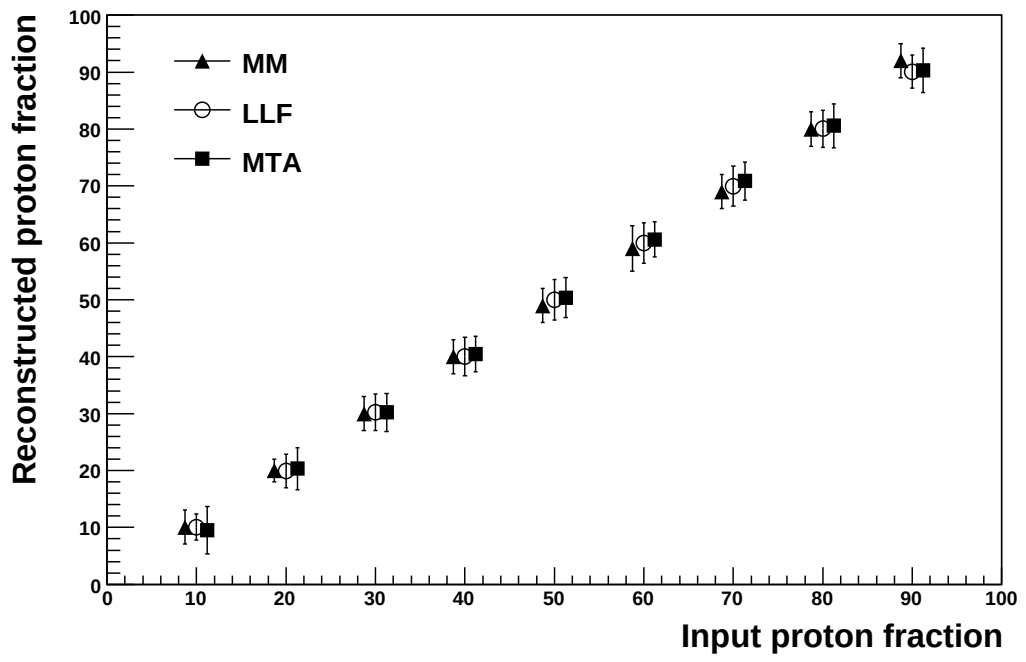

Fig. 3. Reconstructed primary fractions with the MM (full triangles), LLF (empty circles) and MTA (full squares) for different mixtures of protons and iron CONEX showers, with QGSJETII-03, fully simulated through the Auger detector.

## VII. Mean $X_{max}$ estimation from the mass composition analysis

The aim of the described techinques is to derive directly the primary composition of the observed cosmic ray flux. Of course, the obtained primary fractions depend on the hadronic interaction model adopted. Since the study of all the systematics introduced by the models and shower simulation and reconstruction are still under way, we don't report any composition results at this stage. We limit ourselves at checking the consistency of the composition obtained by the different approaches and their ability to reproduce mass sensitive observables.

The change of the mean $X_{max}$ with energy (elongation rate) depends on the primary composition and it is measured directly from fluorescence detectors as at the Auger Observatory. From the primary fractions obtained by the mass composition methods, one can easily derive the mean $X_{max}$ corresponding to the reconstructed mixture. The comparison allows to test if the mass analyses reproduce all the measured elongation rate structures and to have an independent cross-check of the effectiveness of the anti-bias cuts discussed in [2] and [3].

All the hybrid data collected by the Auger Observatory between 1st of December 2004 and the 30th of April 2007 (reported in [2]) have been analyzed with the described mass composition techniques. The studies have been done with a large sample of CONEX simulated showers of protons and iron nuclei, produced with QGSJETII-03. The set has been processed with the Auger analysis framework taking into account the detector evolution with time and the exact working conditions, as done by the Auger Collaboration to compute the hybrid exposure of the Auger Observatory [11]. The analysis has been done above $10^{18}$ eV as the hybrid detector trigger (an Fd event in coincidence with at least one SD station) becomes full efficient both for protons and irons primaries [11].

In each energy bin, the mean $X_{max}$ is given by

$$\langle X_{max} \rangle = P_p \langle X_p \rangle + (1 - P_p) \langle X_{Fe} \rangle \qquad (11)$$

where $P_p$ is the reconstructed proton fraction, while $\langle X_p \rangle$ and $\langle X_{Fe} \rangle$ are the expected mean $X_{max}$ values for proton and iron nuclei. In Fig. 4 the mean $X_{max}$ as a function of energy obtained from the composition results of LLF and MTA, in terms of proton and iron fraction, is shown along with the measured curve [2]. The elongation rates estimated with the two techniques are in agreement with the measured one. All the observed features are well reproduced.

The reconstructed primary fractions obtained by a Monte Carlo based composition analysis are model dependent. To test if the methods could describe the measured elongation rate independently from the hadronic model used, the hybrid data set has been analyzed with a second set of CONEX simulated showers of proton and iron nuclei produced with Sibyll2.1 [12]. The mean $X_{max}$ as a function of energy derived from MTA, in terms of proton and iron fraction with Sibyll2.1 (empty squares) is shown in Fig. 5 along with that one obtained with QGSJETII-03 (empty triangles). The change in the reconstructed primary fractions, due to a different hadronic model, is completely compensated by a change in the expected mean $X_{max}$, giving a compatible curve.

## VIII. Conclusions

Typical performances of the techniques have been evaluated on Monte Carlo data. The input composition abundances are very well reproduced in all the cases.

The techniques have the great advantage to be not biased by the set of analysis cuts applied for the efficiency correction effects, then we can avoid very strong cuts and exploit a larger statistics.

The mass composition methods give primary consistent fractions that allow to reproduce the measured elongation rate reported by the Auger Collaboration at the ICRC 2007, independently from the hadronic model and from the applied set of analysis cuts. The comparison confirms the published Auger results with independent Monte Carlo techniques.

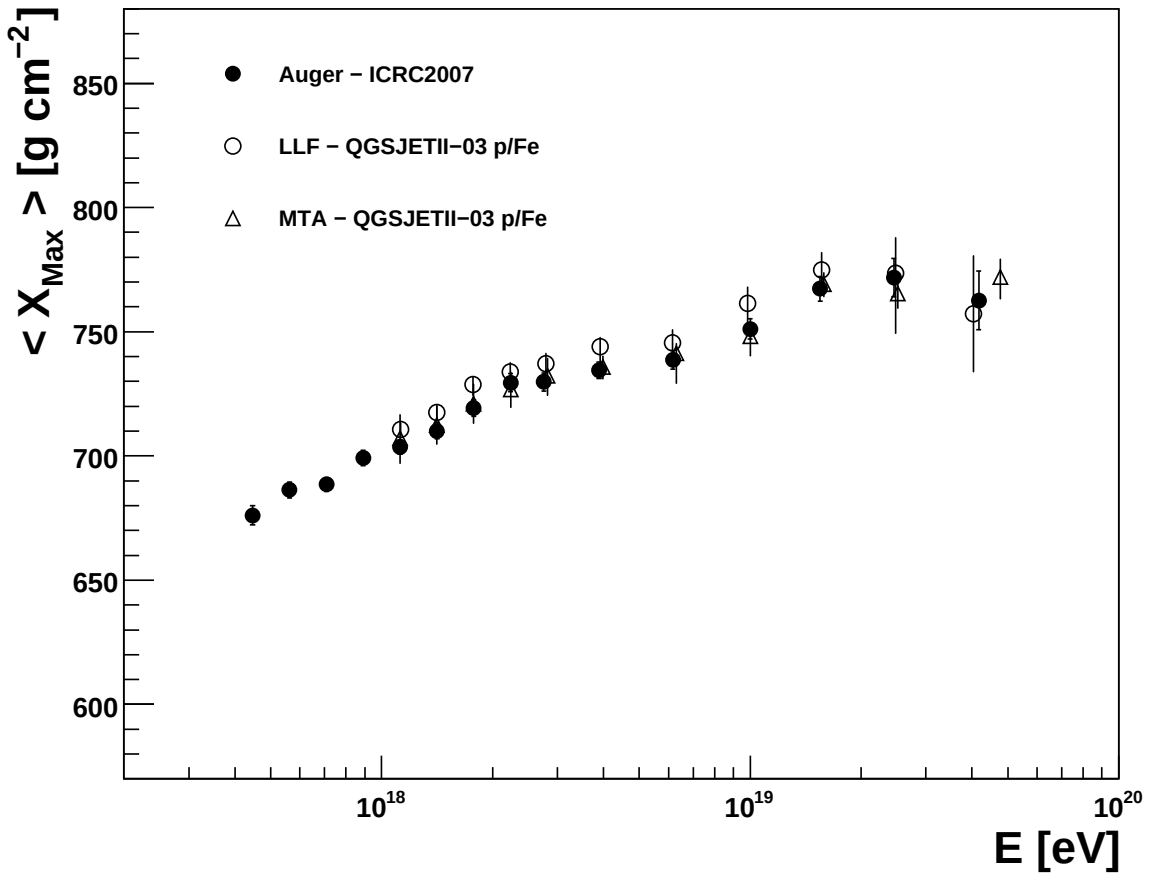

Fig. 4. Mean $X_{max}$ as a function of the energy estimated from LLF (empty circles) and MTA (empty triangles) composition results, obtained using QGJETII-03, compared with that one measured by the Auger Collaboration [2].

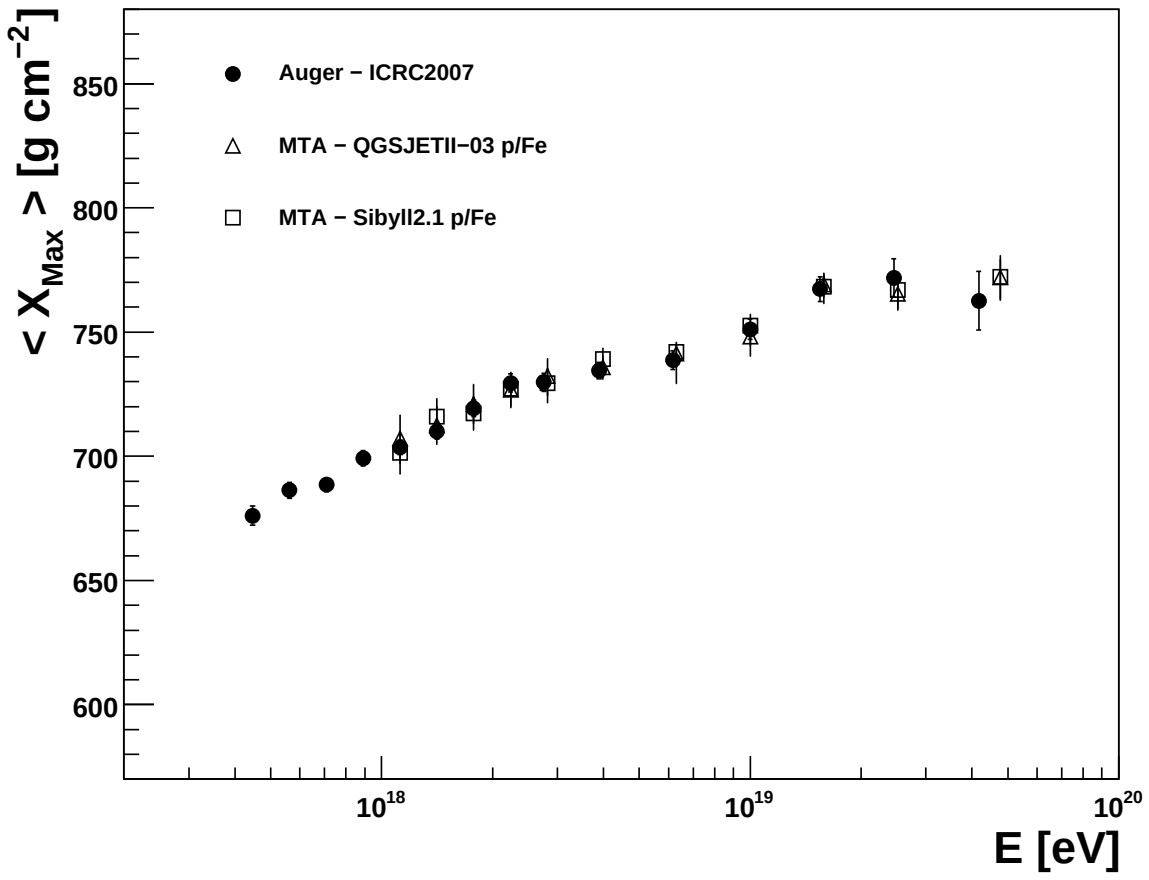

Fig. 5. Mean $X_{max}$ as a function of the energy estimated from MTA primary fractions two different hadronic models: QGSJETII-03 (empty triangles) and Sibyll2.1 (empty squares).

# A Study of the Shower Front in Inclined Showers at the Pierre Auger Observatory

**L. Cazon***, **for the Pierre Auger Collaboration**†

*The Kavli Institute for Cosmological Physics, The University of Chicago, 5801 South Ellis Avenue Chicago, Illinois 60637
†Pierre Auger Observatory, Av San Martin Norte 304,(5613) Malargue, Argentina

***Abstract*. Using a sub-sample of high quality events at zenith angles above 60 degrees the delays in the start-time of the signals detected with water-Cherenkov detectors of the Pierre Auger Observatory with respect to a plane front are compared to those from a model for the arrival time distribution of muons. Good agreement is found and the model correctly accounts for the start-time dependence on the number of particles on each water-Cherenkov detector and on the asymmetries of the shower front. The arrival direction of inclined showers reconstructed by using this model are in good agreement with those obtained with the standard Auger reconstruction.**

## I. Introduction

The southern site of the Pierre Auger Observatory [1] uses 1600 water-Cherenkov detectors, each with an area of 10 m$^2$, and spread over 3000 km$^2$ to collect the secondary particles of extensive air showers. The Cherenkov light is detected by three photomultipliers (PMTs) and the signal is digitalized and recorded as a function of time in 25 ns bins by means of Flash Analog Digital Converters (FADC) whereas conventional GPS receivers are used to synchronize the detectors across the array. The total signal in each detector is measured in Vertical Equivalent Muons (VEM). The time distributions of the signal contain valuable information concerning the arrival direction of the cosmic ray, the longitudinal evolution of the shower and the composition of the primary.

The shower front, defined as the surface containing the first particles to arrive at ground, is estimated by using the onset of the signal in the surface-detector stations, the so called start-time. By fitting a model of the shower front to the experimental start-time, the arrival direction of the air shower can be obtained. The precision achieved in the arrival direction depends on the precision of the detectors clock, the uncertainty in the GPS synchronization, and on the fluctuations in the arrival time of the first detected particles. The standard angular reconstruction uses a model [2] to obtain the start-time variance which is parametrized as a function of the width of the FADC trace and total signal at each triggered station.

A model published elsewhere [3] and updated in [4] describes the arrival time distribution of muons in extensive air showers. Applications of this model span from fast Monte-Carlo simulations used on the Hybrid reconstruction [5] to the reconstruction of longitudinal development of muons based on the surface-detector data only [6]. In this article we use this model to estimate a shower front and compare it to the measured shower front for showers with zenith angles above 60 degrees, which are dominated by muons.

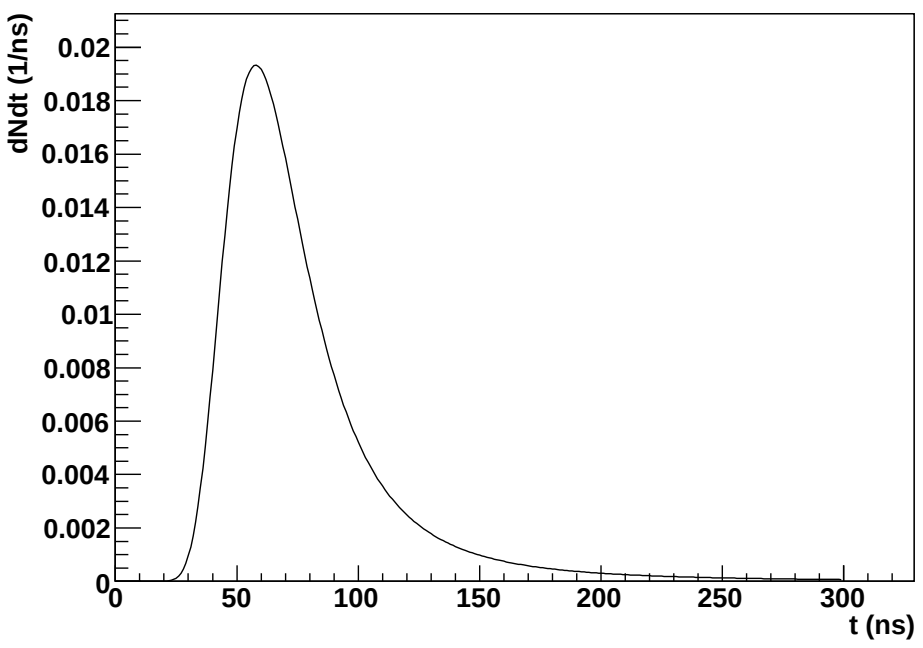

Fig. 1. Example of the probability distribution of the muon arrival times for showers of 70 degrees, at 1000 m from the core.

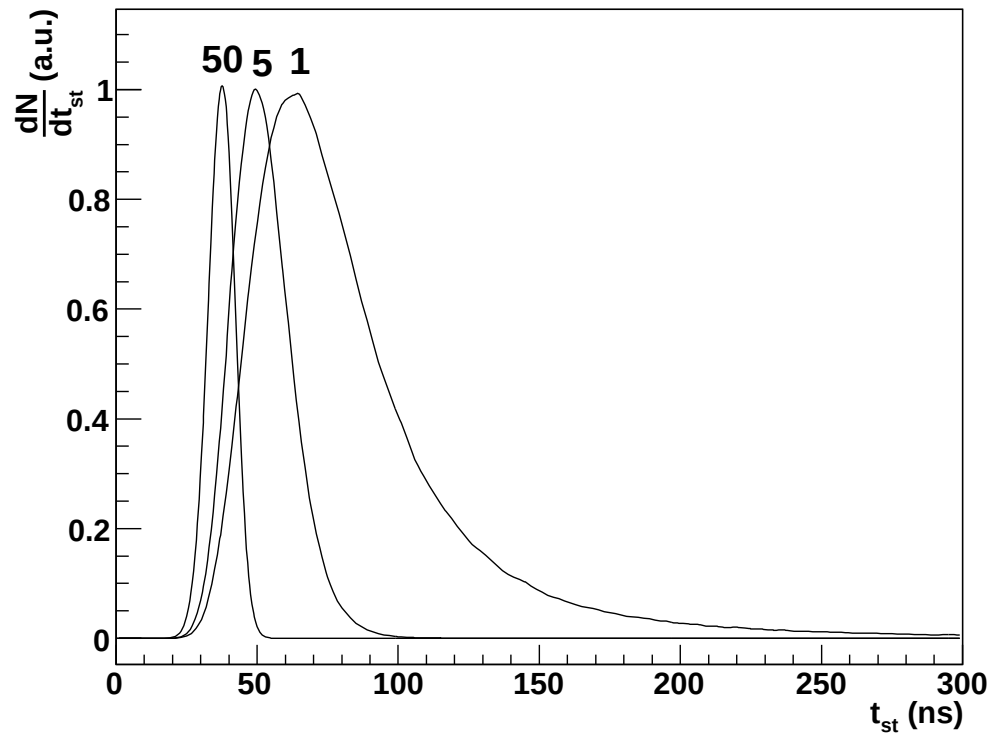

Fig. 2. Example of probability distribution of the start-time for showers of 70 degrees and observed at 1000 m from the core and for $n$=1, $n$=5, $n$=50 muons as labeled.

The model assumes that muons are produced in a narrow cylinder of a few tens of meters around the shower axis. Muons travel practically in straight lines from the production site to the observation point. The effect of multiple scattering and bremstrahlung on their time delay is negligible compared to other effects. From pure geometry, the time delay with respect to a plane front can be easily calculated. This *geometrical delay* establishes a one-to-one correspondence between the production distance $z$, measured along the shower axis from the ground up to the production site, and the arrival time delay $t$, which is the time elapsed from the arrival time of the shower front plane and the arrival time of the particle. This one-to-one correspondence is different for each relative position with respect to the shower axis. If the muons are produced at distances $z$ with a distribution $\frac{dN}{dz}$, the corresponding arrival time distribution is simply[1] $\frac{dN}{dt} = \frac{dN}{dz}\frac{dz}{dt}$. The distribution $\frac{dN}{dz}$ depends on composition, zenith angle, energy of the primary and it is affected by shower-to-shower fluctuations. The model for this article uses a parametrization of the average $\frac{dN}{dz}$ of $10^{19}$ eV proton showers simulated with AIRES 2.6.0 [7] QGSJET01 [8].

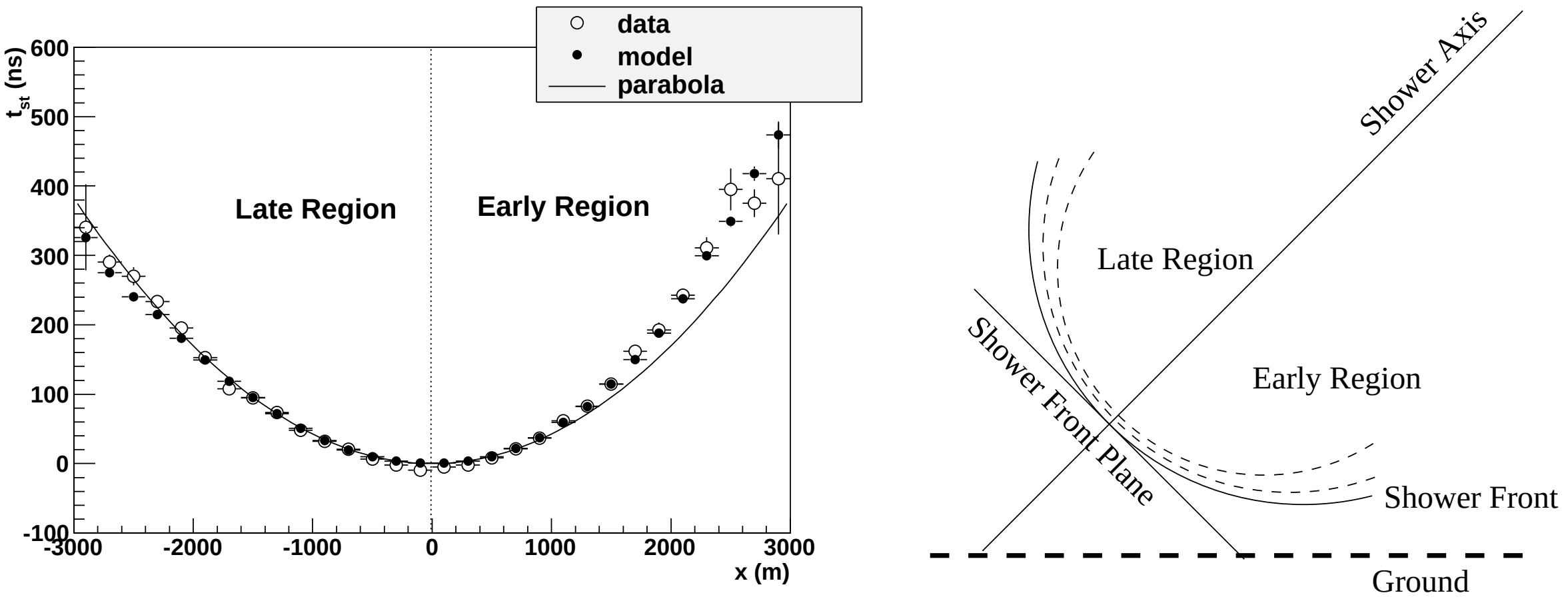

(a) Average Shower Front (b) Schematic View

Fig. 3. (a) Average start-times for showers at 70 degrees for data (open circles) and model (solid circles). Also shown a parabola. (b) Schematic view of the geometry of the shower pancake. The shower front is just the surface containing the first particles arriving to ground. Also labeled the early and late regions.

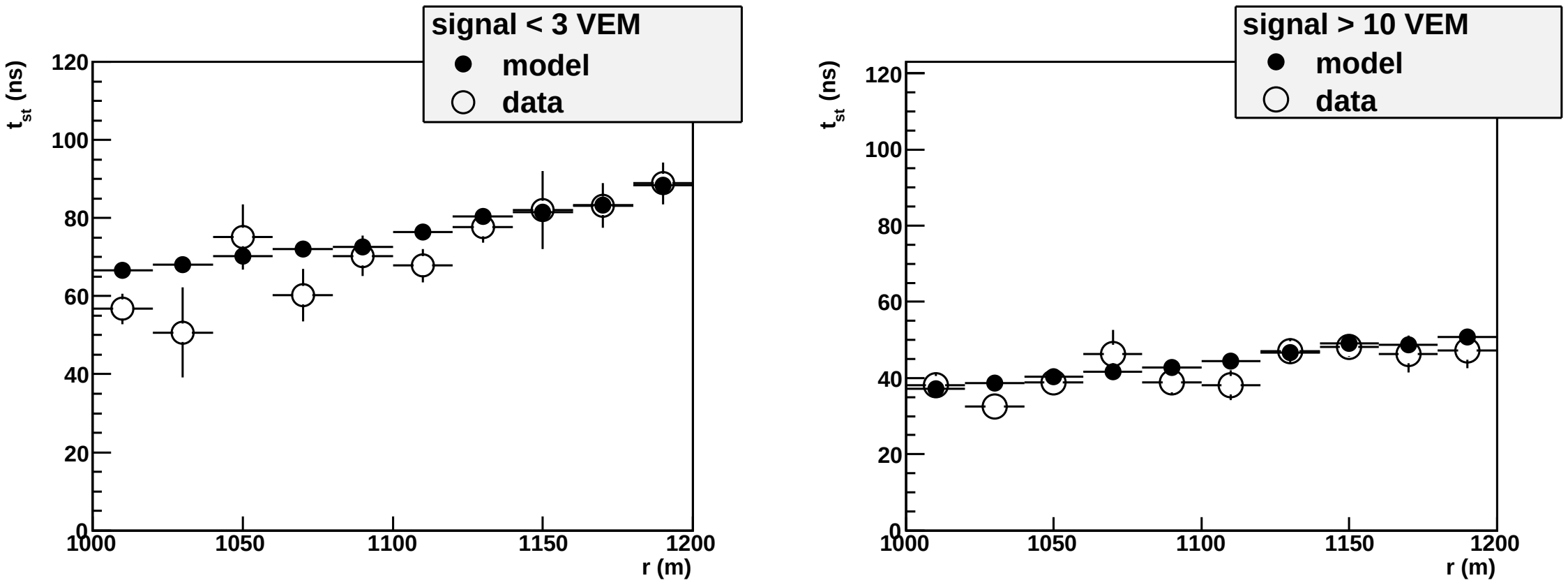

Fig. 4. Comparison of average start-times for showers at 70 degrees at a given range of distances to the shower core in two very different conditions of signal at the station, namely less than 3 VEM (left panel), and more than 10 VEM (right panel).

The description provided by this model has been verified with detailed Monte Carlo simulations [3][4]. Figure 1 displays an example of the muon arrival time distribution predicted by this model, $\frac{dN}{dt}$, for showers of 70 degrees at 1000 m from the core.

## II. The start-time

The expected start-time is calculated from the model at each station in the following way:

- First, the signal produced solely by muons is estimated by taking the VEM content at the detector and subtracting a parametrized EM component [9]. Muons arrive earlier than the electromagnetic component, which is affected by multiple scattering, and is of the order of 15%-20% of the total signal in inclined showers.

[1]There is an additional source of delay coming from the fact that muons have finite energies and they do not travel at the speed of light. This is the so called *kinematic delay* and it is accounted for by modeling the energy spectrum of muons at the observation point. At large distances from the core the *geometrical delay* dominates, and the *kinematic delay* is just a small correction. Besides, the effects the geomagnetic field are negligible, and only for showers with $\theta > 85$ deg the arrival time distributions start to show some visible effect due to the extra path of the bent muon trajectories.

- The number of muons $n$ is calculated dividing the remaining signal by the average signal produced by muons, which is proportional [2] to the average tracklength of the muon in the detector: $n \simeq S\frac{h}{L_\alpha}$, where $S$ is the VEM content of the station, $h$ is the height of water in the the detector, and $L_\alpha$ is the average tracklength at a given incident angle to the station $\alpha$, which can be approximated by the zenith angle of the shower, becoming $\alpha = \theta$.
- Once the $\frac{dN}{dt}$ distribution is calculated for a given zenith angle and the relative position of the station respect to the core, (see Figure 1 for an example of $\theta = 70$ deg and $r = 1000$ m), it is sampled $n$ times simulating the physical realisation of $n$ muons arriving at ground. The first or earliest muon out of $n$ is kept, and its time $t_{st}$ becomes a possible start-time realisation. Figure 2 displays the distribution of the start-times, $\frac{dN}{dt_{st}}$, for n=1, n=5, and n=50 muons.

Note that we cannot predict $t_{st}$ in each single realisation of real data, but only its distribution. The average $< t_{st} >$ becomes the expected start-time predicted by the model, and the RMS of this distribution, $\sigma_{st}$, corresponds to the expected typical fluctuation or uncertainty of any $t_{st}$. The value of $\sigma_{st}$ can be extremely small because the arrival time distribution of particles can be very narrow when the station is near the core, or when the number of muons $n$ is large. To account for the finite resolution of the detection device in the final start-time uncertainty, a constant term $b$ has been introduced. The total uncertainty therefore becomes

$$\sigma_{det} = \sqrt{\sigma_{st}^2 + b^2} \quad (1)$$

Pairs of adjacent stations separated by 11 m were used to adjust the value of $b$ and make $\sigma_{det}$ equal to the average start-time fluctuations of data. The value that best meets this requirement for a broad range of signal at each station and distance to the core is $b = 20$ ns. The GPS time resolution [10] ($\sim 10$ ns) and the procedure of signal digitization in the 25 ns bins accounts for 12 ns, which is not enough to explain the value of $b$. A possible explanation is that $b$ absorbs a mismatch of the model with respect to data. The model for the start-times uncertainties described in [2] which is used in the standard reconstruction, finds a value for $b$ in better agreement to the expectations by not attempting to describe the details of the shower front itself and using an estimation of the overall length of the arrival time distribution of the particles from the data recorded in each station.

## III. Validation

Data collected from January 2004 through December 2008 were reconstructed using the horizontal reconstruction chain [11], and events satisfying the T5 quality trigger were selected [12]. The described time model needs the incoming direction of the shower and also the core position, which are obtained from an iterative angular-core reconstruction [12].

Figure 3a displays the average of the predicted and observed start-times for a subset of showers of 70 degrees as a function of the perpendicular distance to the shower core. The overall curvature features are well described by the model, although data tend to show slightly more curvature than the model. A parabola is shown, to illustrate the asymmetries on the arrival times between the early and late region, which are depicted in Figure 3b. This asymmetry is naturally accounted for by the model through the different distances to the production site along the shower axis.

Figure 4 displays a comparison of the average start-times at a given range of distances to the shower core in two very different conditions of signal at the station, namely less than 3 VEM (left panel), and more than 10 VEM (right panel). One can see that the effect on the start-time produced by the different sampling of the number of muons is to bring the start-time to earlier times when $n$ is larger.

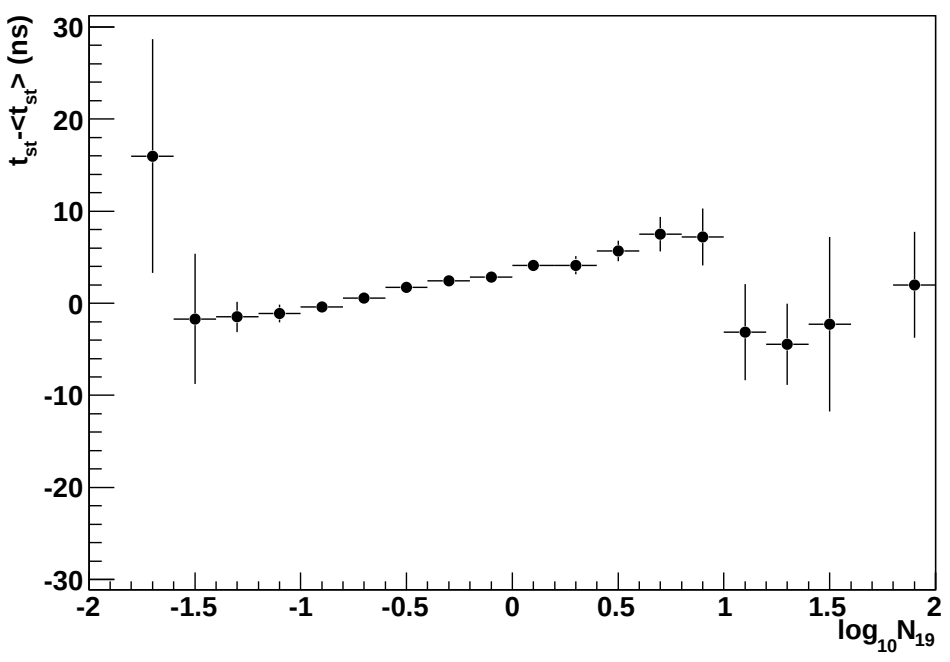

Fig. 5. Average value of the difference between $t_{st}$ (data) and $< t_{st} >$ (model) as a function of the shower size parameter $N_{19}$.

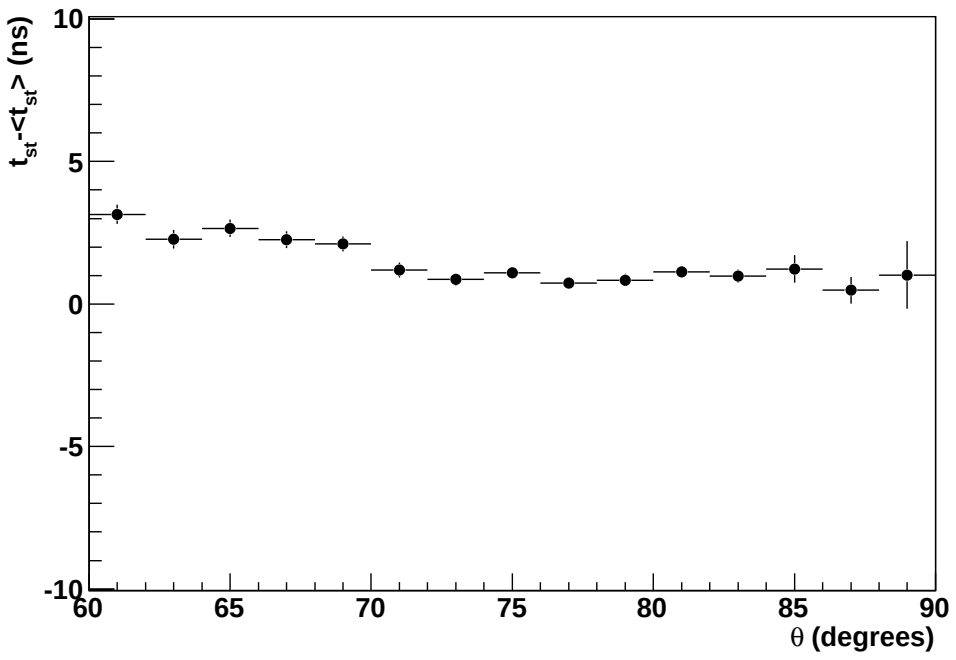

Fig. 6. Average value of the difference between $t_{st}$ (data) and $< t_{st} >$ (model) as a function of the zenith angle $\theta$.

Figure 5 and 6 display the average value of $t_{st} - < t_{st} >$ as a function of the zenith angle and of the shower size parameter $N_{19}$, which is approximately proportional to the energy of the primary and it is used in the inclined showers reconstruction [11]. The average typical differences between the start-time given by the model and data are below 10 ns within this range of $N_{19}$ and zenith angles.

[2] A more sophisticated approach that takes into account the distribution of signal $S$ for each number of muons $n$ is currently under study. Note that $S$ is affected by the muon energy spectrum and the distribution of tracklengths $L_\alpha$.

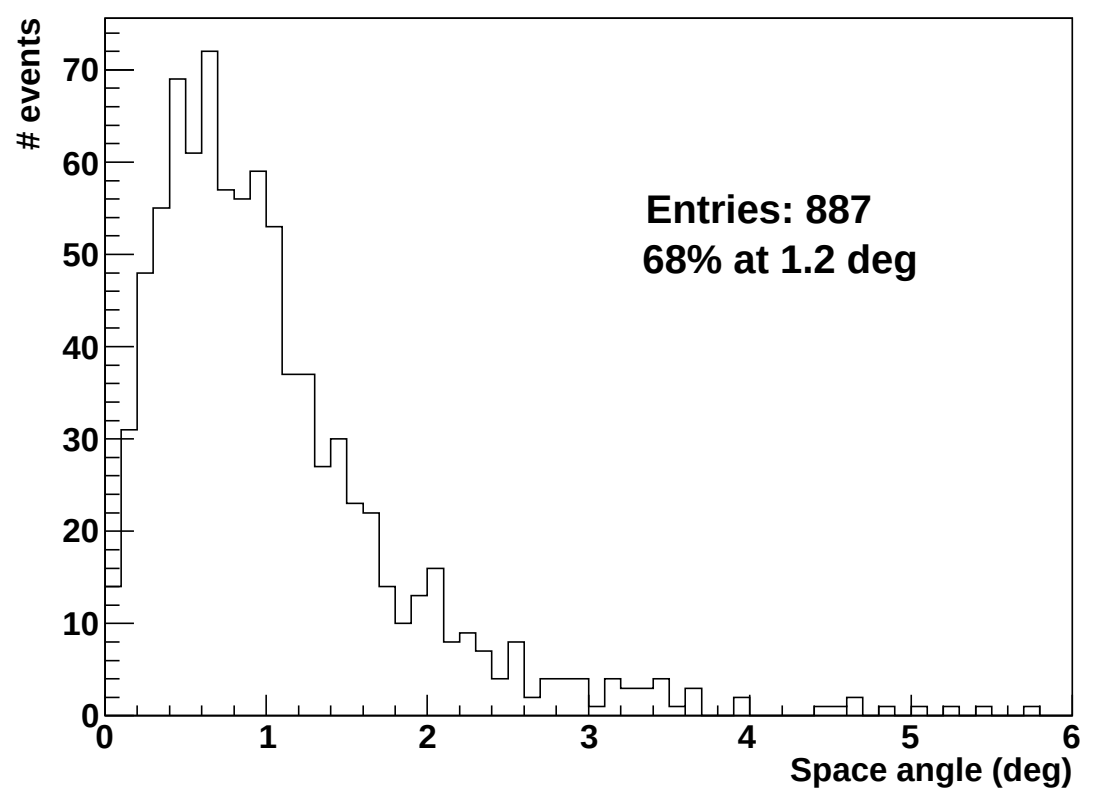

Fig. 7. Comparison of the angular reconstruction using the present model to the Hybrid Auger reconstruction.

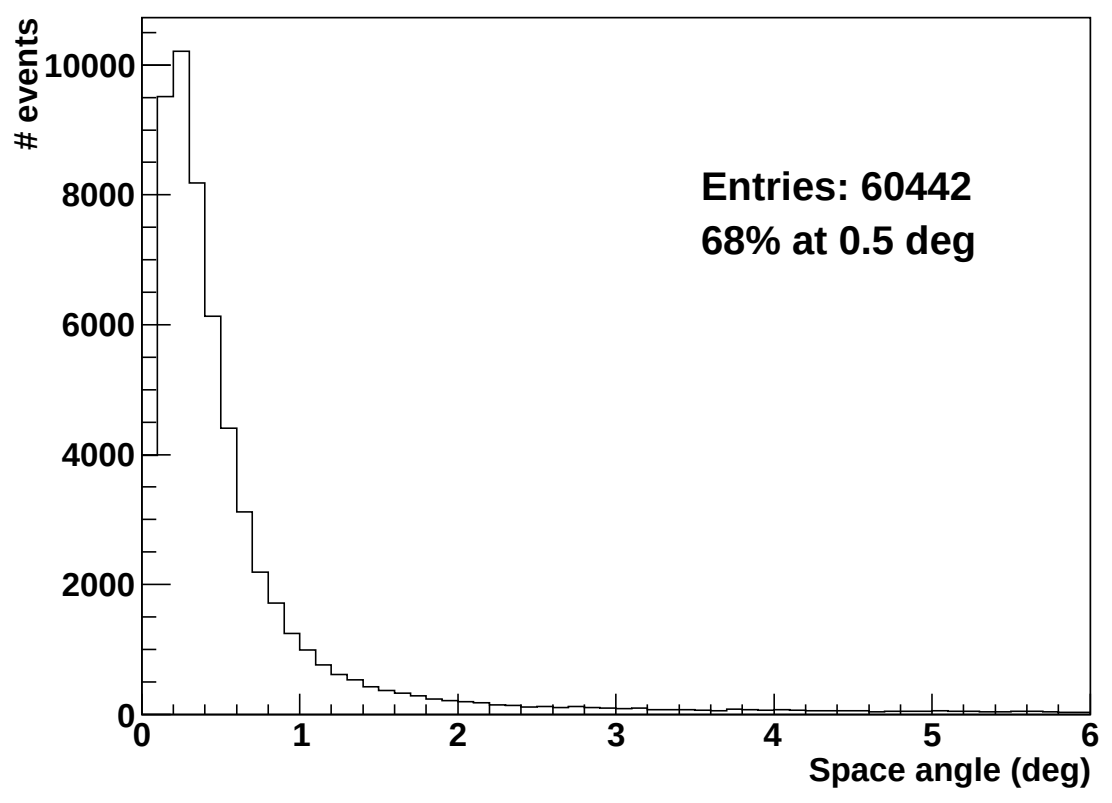

Fig. 8. Comparison of the angular reconstruction using the present model to the one currently used in the Auger reconstruction chain.

It is expected that further refinements of the model, like a parametrization of the average $\frac{dN}{dz}$ distribution from real data [6], or accounting for the shower to shower fluctuations may further improve the results.

An independent indication that the model is describing the data reasonably well comes from a comparison of the angular reconstruction that uses the current prediction shower front to that coming from the hybrid reconstruction, which is shown in Figure 7. 68% of events have a space angle difference less than 1.2 degrees, which result in an angular resolution of $\sim 1$ degrees. (The hybrid resolution is $\sim 0.6$ degrees). Figure 8 displays the comparison between the angular reconstruction that uses the current model with the standard Auger reconstruction: 68% of the events have a space angle difference less than 0.5 degrees, which is compatible with the angular resolution of Auger [13].

## IV. Conclusions

A model for the arrival time distribution of muons has been used to predict the shower front. This model uses as input a parametrization of the muon production distance distribution $\frac{dN}{dz}$. The model accounts for the different curvatures due to the early-late asymmetries of the shower front, and the number of particles detected at the station. The results from the model have been compared to data coming from real events above 60 degrees showing good agreement. When the predicted shower front is used to reconstruct the arrival directions, results are consistent with those coming from the standard reconstruction and hybrid reconstruction. Future improvements of the model include the use of $\frac{dN}{dz}$ distributions deduced from data.

# UHE neutrino signatures in the surface detector of the Pierre Auger Observatory

**D. Góra*† for the Pierre Auger Collaboration‡**
*Karlsruhe Institute of Technology (KIT), D-76021 Karlsruhe, Germany
†Institute of Nuclear Physics PAN, ul. Radzikowskiego 152, 31-342 Kraków, Poland
‡Av. San Martín Norte 304 (5613) Malargüe, Prov. de Mendoza, Argentina

***Abstract*. The Pierre Auger Observatory has the capability of detecting ultra-high energy neutrinos. The method adopted is to search for very inclined young showers. The properties of such showers that start deep in the atmosphere are very different at ground level from those of showers initiated in the upper atmosphere by protons or nuclei. The neutrino events would have a significant electromagnetic component leading to a broad time structure of detected signals in contrast to nucleonic-induced showers. In this paper we present several observables that are being used to identify neutrino candidates and show that the configuration of the surface detectors of the Auger Observatory has a satisfactory discrimination power against the larger background of nucleonic showers over a broad angular range.**

## I. INTRODUCTION

The detection of ultra high energy (UHE) cosmic neutrinos, above $10^{18}$ eV, is important as it may allow us to identify the most powerful sources of cosmic rays (CR) in the Universe. Essentially all models of UHECRs production predict neutrinos as a result of the decay of charged pions produced in interactions of cosmic rays within the sources themselves or while propagating through background radiation fields [1]. For example, UHECR protons interacting with the cosmic microwave background (CMB) give rise to the so called "cosmogenic" or GZK neutrinos [2]. The cosmogenic neutrino flux is somewhat uncertain since it depends on the primary UHECR composition and on the nature and cosmological evolution of the sources as well as on their spatial distribution [3]. In general, about 1% of cosmogenic neutrinos from the ultra-high energy cosmic ray flux is expected.

Due to their low interaction probability, neutrinos need to interact with a large amount of matter to be detected. One of the detection techniques is based on the observation of extensive air showers (EAS) in the atmosphere. In the atmosphere so-called down-going neutrinos of all flavours interacting through charge or neutral currents can produce EAS potentially detectable by a large ground detector such as the Pierre Auger Observatory [4]. When propagating through the Earth only tau neutrinos skimming the Earth and producing an emerging tau lepton which decays in flight may initiate detectable air showers above the ground [5], [6].

One of the experimental challenges is to discriminate neutrino-induced showers from the background of showers initiated by UHECRs. The underlying concept of neutrino identification is rather straightforward. Whereas proton or nuclei and photons interact shortly after having entered the atmosphere, neutrinos may penetrate a large amount of matter undisturbed and generate showers close to the surface array. The differences between showers developing close to the detector – so-called young showers – and showers interacting early in the atmosphere – old showers – becomes more and more pronounced as we consider larger angles of incidence. In case of showers initiated by protons and nuclei, which interact soon after entering the atmosphere, only high-energy muons can survive at high zenith angles. As a result, the detected showers show a thin and flat front which leads to short detected signals ($\sim$ 100 ns). In case of young neutrino-induced showers a significant electromagnetic component (EM) is present at the ground as well. The shower front is curved and thick and leads to broad signals, lasting up to a few microseconds.

With the surface detector array (SD) of the Auger Observatory, which consists of 1600 water Cherenkov detectors with 1.5 km spacing, we can identify young showers because the signal in each tank is digitized with 25 ns time resolution, allowing us to distinguish the narrow signals in time expected from old showers, from the broad signals expected from a young shower.

In this contribution, we present the criteria used to identify neutrino-induced showers, the important observables, the neutrino identification efficiencies, and the procedure to simulate neutrino induced showers.

## II. "EARTH-SKIMMING" TAU NEUTRINOS

The SD detector of the Auger Observatory is sensitive to Earth-skimming tau neutrinos [7], [8], [9]. These are expected to be observed by detecting showers induced by the decay of emerging $\tau$ leptons, after the propagation of of $\nu_\tau$s through the Earth, see Fig. 1 (upper panel). The first step towards identification of $\nu_\tau$ induced showers consists of selecting very inclined showers that have most of the stations with signals sufficiently spread in time. Young showers are expected to trigger detector stations with broad signals releasing a so-called 'Time Over Threshold' (ToT) trigger [7]. Counting ToTs stations can help identifying young showers. At this stage also a cut of the area of the signal over its peak (AoP)[1] value is applied to reject ToT local triggers produced by consecutive muons hitting a station. Then the elongation of footprint, defined by the ratio of length (L) over width (W) of the shower pattern on ground, and the mean apparent velocity, are basic ingredients to identify very inclined showers [7], see Fig. 1 (bottom panel) for the explanation of these observables.

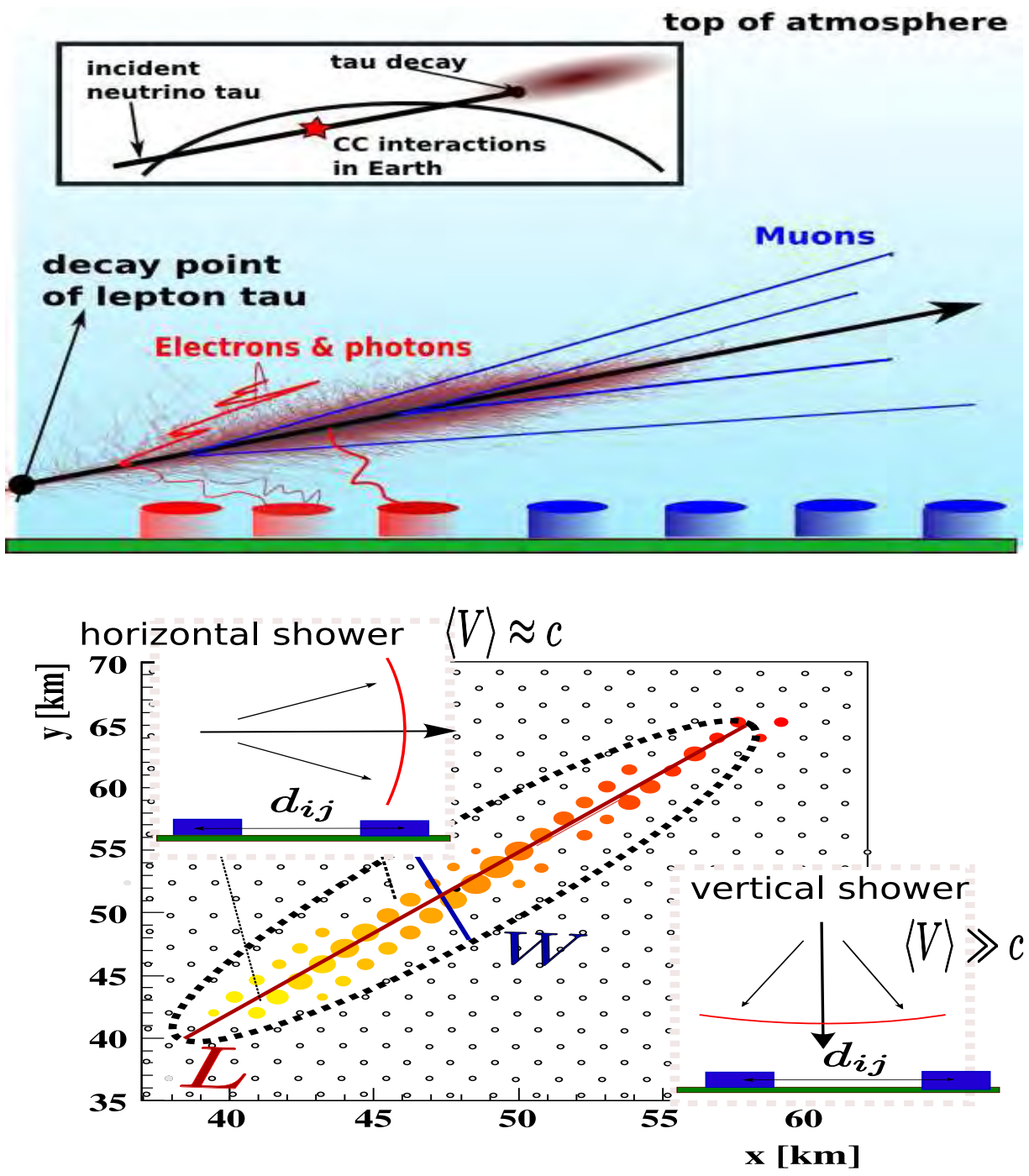

Fig. 1. (Upper panel) The sketch of a shower induced by the decay of a $\tau$ lepton emerging from the Earth after originating from an Earth-skimming $\nu_\tau$. The earliest stations are mostly triggered by electrons and $\gamma$s; (bottom panel) sketch of length $(L)$ over width $(W)$ of a footprint and determination of the apparent velocity $(\langle V\rangle)$. The $\langle V\rangle$ is given by averaging the apparent velocity, $v_{ij} = d_{ij}/\Delta t_{ij}$ where $d_{ij}$ is the distance between couples of stations, projected onto the direction defined by the length of the footprint, $L$, and $\Delta t_{ij}$ the difference in their signal start times.

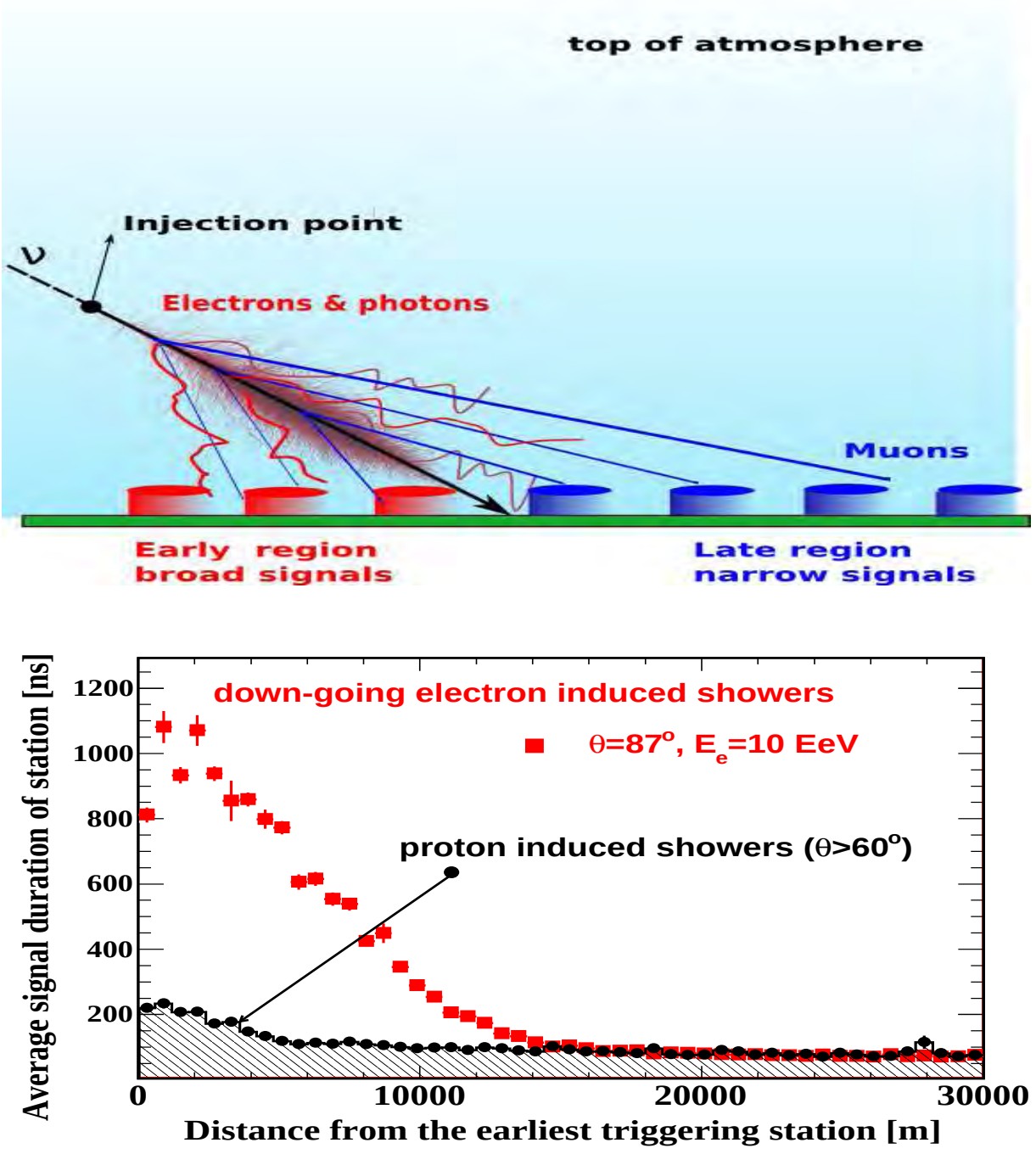

Fig. 2. (Upper panel) Sketch of a down-going shower initiated in the interaction of a $\nu$ in the atmosphere close to the ground; In the "early" ("late") region of the shower before (after) the shower axis hits the ground we expect broad (narrow) signals in time due to electromagnetic (muonic) component of the shower; (bottom panel) the average signal duration of the station as a function of the distance from the earliest triggering station.

The mean apparent velocity, $\langle V\rangle$ is expected to be compatible with the speed of light for quasi-horizontal showers within its statistical uncertainty $\sigma_{\langle V\rangle}$ [8]. Finally compact configurations of selected ToTs complete the expected picture of young $\nu_\tau$-induced shower footprints. These criteria were used to calculate an upper limit on the diffuse flux UHE $\nu_\tau$ [8] with the Auger Observatory and an update of this limit [9], [10].

## III. "DOWN-GOING" NEUTRINOS

The SD array is also sensitive to neutrinos interacting in the atmosphere and inducing showers close to the ground [11], [12]. Down-going neutrinos of any flavours may interact through both charged (CC) and neutral current (NC) interactions producing hadronic and/or electromagnetic showers. In case of $\nu_e$ CC interactions, the resulting electrons are expected to induce EM showers at the same point where hadronic products induce a hadronic shower. In this case the CC reaction are simulated in detail using HERWIG Monte Carlo event generator [13]. HERWIG is an event generator for high-energy processes, including the simulation of hadronic final states and the internal jet structure. The hadronic showers induced by outgoing hadrons are practically indistinguishable in case of $\nu$ NC interactions, so they are simulated in the same way for three neutrino flavours. In case of $\nu_\mu$ CC interactions the produced muon is expected to induce shower which are generally weaker i.e. with a smaller energy transfer to the EAS, and thus with suppressed longitudinal profile and much fewer particles on ground. As a consequence, the detection probability of such shower is low and therefore the produced muon is neglected and only the hadronic component is simulated with the same procedure adopted for $\nu$ NC interactions. In case of down-going $\nu_\tau$ the produced $\tau$ lepton can travel some distance in the atmosphere, and then decay into particle which can induce a detectable shower. Thus, the outcoming hadronic showers initiated by $\nu_\tau$ interactions are usually separated by a certain distance from the shower initiated by the tau decay. In this particular case, $\tau$ decays were simulated using TAUOLA [16]. The secondary particles produced by HERWIG or TAUOLA are injected into the extensive air shower generator AIRES [17] to produce lateral profiles of the shower development. Shower simulations were performed including the geographic conditions of the site (e.g. geomagnetic field) for different zenith angles $\theta = 75°, 80°, 85°, 87°$, $88°$ and $89°$ and random azimuth angles between $0°$ and $360°$ and different hadronic models: QGSJET II [14] and Sibyll [15]. The secondary particles are injected at different slant depths measured from the ground up to a maximum value depending on $\theta$. Finally the response of the SD array is simulated in detail using the Offline simulation package [18]. In total about 20,000 showers induced by down-going decaying $\tau$ leptons were simulated and about 36,000 events for electron induced showers. These neutrino simulations were used to estimate the expected neutrino signal and efficiency of detection of the neutrinos.

[1]The peak corresponds to the maximum measured current of recorded trace at a single water-Cherenkov detector.

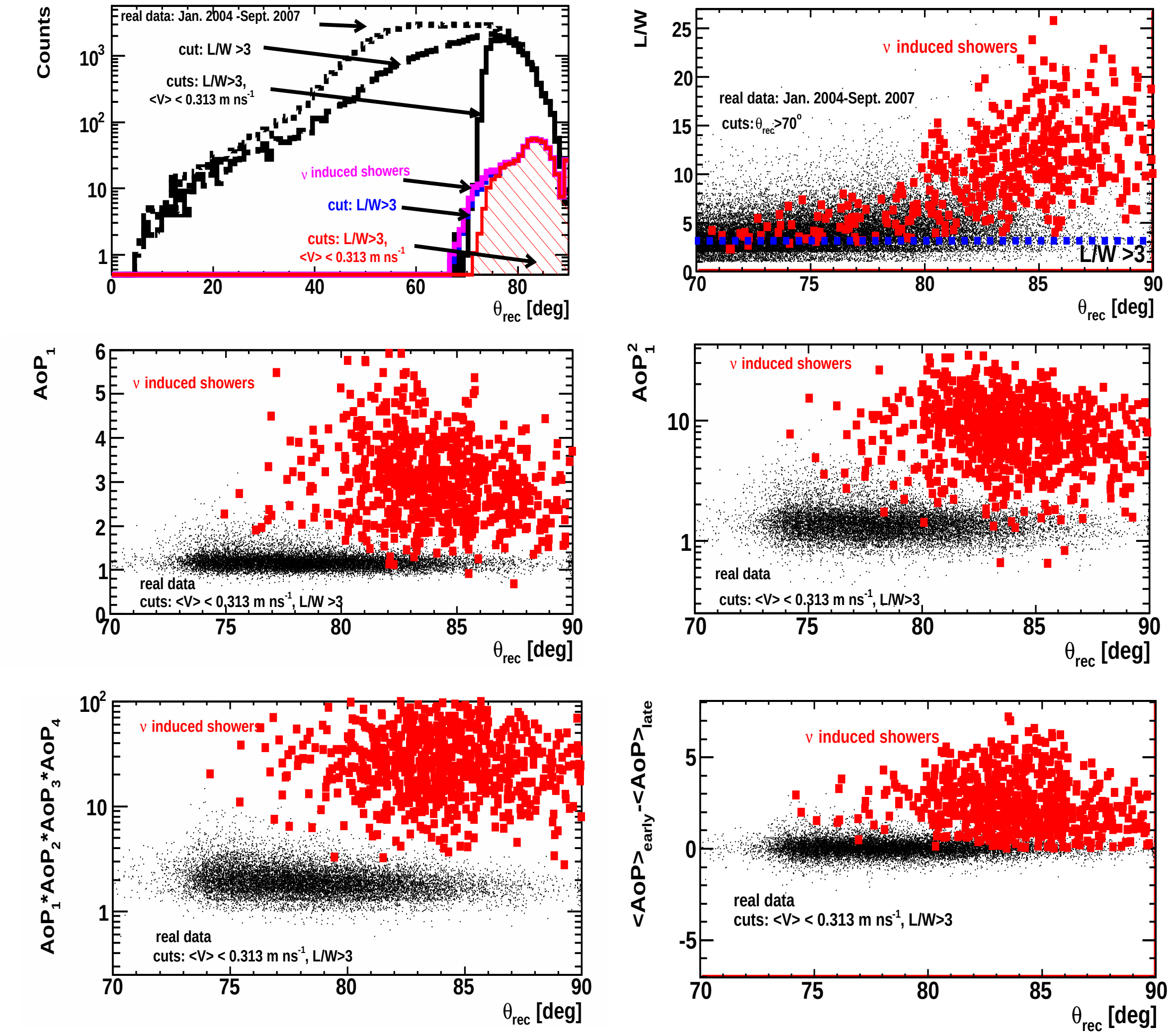

Fig. 3. (Left upper panel) The zenith angle distribution of neutrinos with $E^{-2}$ flux and real events; (right upper panel) the ratio $L/W$ as the function of the reconstructed zenith angle. Neutrino induced showers have larger ratio $L/W$ than real data at high zenith angles. The area over peak for first triggering station ($\text{AoP}_1$) (left middle panel) the square of the area over peak for first triggering station ($\text{AoP}_1^2$) (right middle panel), the product of AoP of four first triggering stations (left bottom panel) and a global early-late asymmetry parameter ($\langle\text{AoP}\rangle_{early} - \langle\text{AoP}\rangle_{late}$) as the function of zenith angle.

The criterion to identify young, inclined, down-going showers consists of looking for broad time signals as in the case of up-going neutrinos, at least in the early region, i.e. in those stations triggered before the shower core hits the ground [12]. The physical basis for this criterion is the large asymmetry in the time spread of signals that one expects for very inclined young showers, in which the late front of the shower typically has to cross a much larger grammage of atmosphere than the early front, and as a consequence suffers more attenuation, see Fig. 2 (upper panel). This has been confirmed by simulations of $\nu$-induced showers as is shown in Fig. 2 (bottom panel). The time signal for $\nu$-showers is expected to be broader around the position of the maximum of the shower development. Broader signals are expected to last about 1000 ns, while the duration decreases to a value of about 150 ns downstream in the latest stations which are hit by the muonic tail of the shower development. For hadronic showers with $\theta > 60°$, the expected duration of the signals is almost constant with an average value of about 150 ns. From Fig. 2 (bottom panel) we can see that a good identification criterion is to require broad signals in the first triggered stations of an event.

In the case of down-going neutrinos the general procedure to extract a neutrino induced shower from real data is similar to the procedure used for Earth-skimming neutrinos, i.e. the inclined events are extracted from real data using the apparent velocity and $L/W$ cut and the criterion for looking for events with broad signal in time are applied. However, there are some differences. The selection criteria cannot be the same as for up-going $\nu_\tau$, because in case of down-going neutrinos we are sensitive for a larger zenith angle range (about $15^\circ$ above the horizon instead about $5^\circ$ below horizon for up-going $\nu_\tau$), which also means a larger background contribution and thus a more demanding selection procedure [10].

In Fig. 3 (left upper panel) the zenith angle distribution of real data and simulated neutrino events is shown. The $\langle V \rangle$ and the ratio $L/W$ cut can extract inclined events from real data, see also Fig. 3 (right upper panel). To extract young showers with broad signals, the area over the peak (AoP) of the first four stations its square ($\mathrm{AoP}^2$), their product ($\mathrm{AoP}_1$*$\mathrm{AoP}_2$*$\mathrm{AoP}_3$*$\mathrm{AoP}_4$) and a global early-late asymmetry parameter of the event $(\langle \mathrm{AoP} \rangle_{early} - \langle \mathrm{AoP} \rangle_{late})^2$ can be used. These observables were used to discriminate neutrino showers by using the Fisher method, see [10] for more details. As an example in Fig. 3 (middle panels) distributions of $\mathrm{AoP}_1$ and $\mathrm{AoP}_1^2$ for the first triggering station are shown. In Fig. 3 (lower panels) we also show the product $\mathrm{AoP}_1$*$\mathrm{AoP}_2$*$\mathrm{AoP}_3$*$\mathrm{AoP}_4$ (left panel) and the global early-late asymmetry parameter $\langle \mathrm{AoP} \rangle_{early} - \langle \mathrm{AoP} \rangle_{late}$ (right panel) for real data and MC simulated neutrinos. The good separation is clearly visible between neutrino simulated showers and measured inclined events. The separation is better at large zenith angles where the background signal (real data events) is less abundant. This example demonstrates that the SD array has a satisfactory discriminating power against the larger background of nucleonic showers at zenith angles larger than about $75^\circ$.

In Fig. 4 the neutrino identification efficiency, $\epsilon$ (the fraction of $\nu$-induced showers triggering SD array and passing the neutrino identification criteria [10]) is shown. It is clear that $\epsilon$ depends on the zenith angle and type of interactions. The efficiency as well as the range of slant depth grows as the zenith angle increases. Only for showers very close to the SD array does it drop dramatically since the shower does not cross sufficient grammage to develop in the direction transverse to the shower axis. The efficiencies for NC are much lower than for CC for the same neutrino energy and zenith angle. This is due to the fact that in NC reactions the fragments of a target nucleus induce a pure hadronic shower with a small fraction (about 20%) of energy transfered to the EAS while in CC $\nu_e$ reaction the rest of the energy goes to an additional EM shower. The identification efficiency depends also on the neutrino flavour due to different energy fractions transferred to the induced shower. In CC $\nu_\tau$ interactions, if the lepton tau decays in flight, only a fraction of its energy is converted into a $\tau$-induced shower. In a $\nu_\mu$ CC interaction, the produced muon induce a shower which is in general weaker, with a small energy transfer to an EAS with very low probability to trigger the SD array. Thus the $\nu_e$ CC induced showers give the main contribution to the expected event rate.

[2]The global early-late asymmetry parameter is defined as the difference between average value of AoPs calculated for the first triggered stations and the last triggered stations of the event. If the number of stations is odd the station in middle is ignored. If the event multiplicity is larger than 8 stations only the first/last four stations are used.

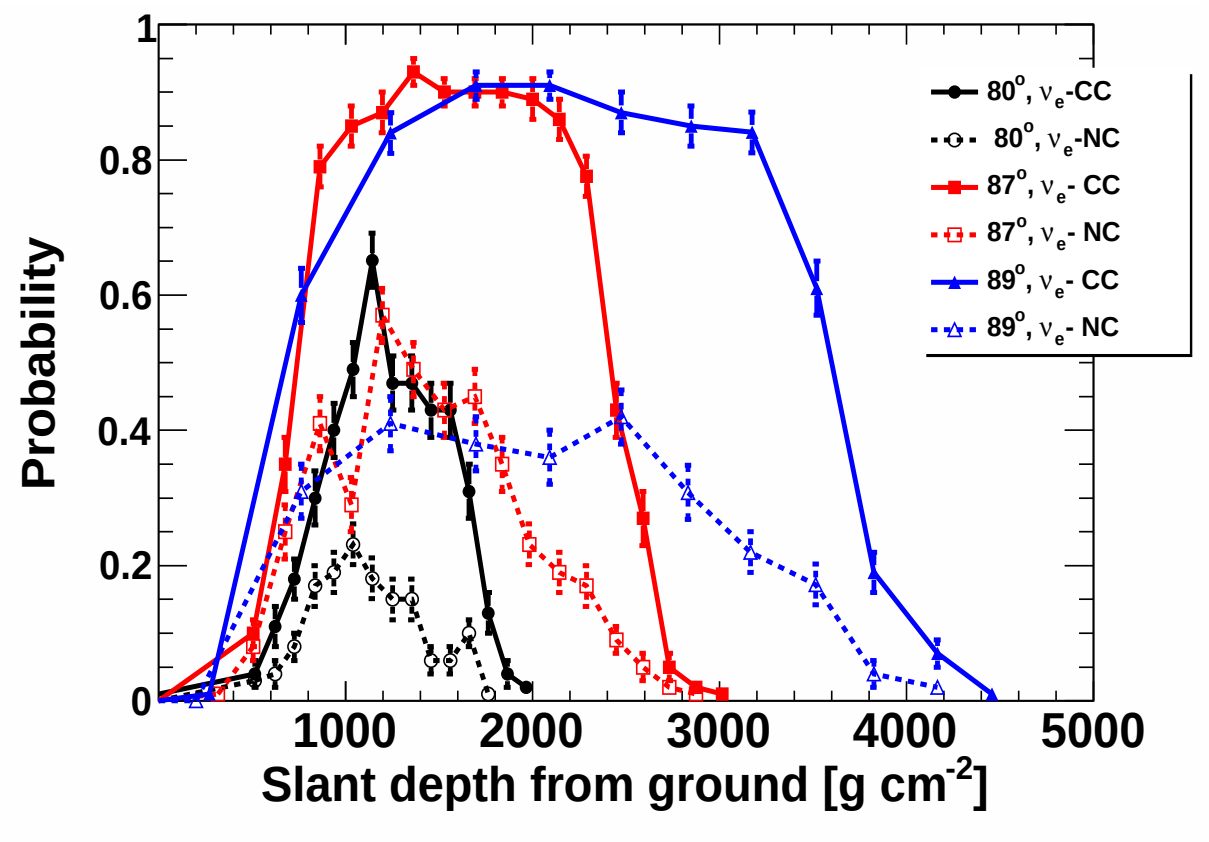

Fig. 4. The $\nu_e$ identification efficiency as a function of the neutrino interaction point for different zenith angle and energy 1 EeV.

## IV. Conclusions

To conclude we have shown that neutrino induced shower can be identified by the SD of the Auger Observatory. The key to for $\nu$ identification is the presence of a significant EM component. By means of Monte Carlo simulations we have identified the parameter space where the efficiency of neutrino identification is significant.

# The electromagnetic component of inclined air showers at the Pierre Auger Observatory

**Inés Valiño* for the Pierre Auger Collaboration†**

*Karlsruhe Institute of Technology, POB 3640, D-76021 Karlsruhe, Germany
†Av. San Martin Norte 304, 5613 Malargüe, Argentina

***Abstract*. Muons, accompanied by secondary electrons, dominate the characteristics of inclined air showers above $60°$. The characteristics of the signal induced by the electromagnetic component in the water-Cherenkov detectors of the Pierre Auger Observatory are studied using Monte Carlo simulations. The relative contributions of the electromagnetic component to the total signal in a detector are characterised as a function of the primary energy, for different assumptions about mass composition of the primary cosmic rays and for different hadronic models.**

## I. INTRODUCTION

Inclined air showers are conventionally defined as those arriving at ground with zenith angles $\theta$ above $60°$. At large zenith angles the electromagnetic (EM) component in air showers, mainly produced by the decay of $\pi^0$s, is largely absorbed in the vastly enhanced atmospheric depth crossed by the shower before reaching ground, so in a first approximation only the more penetrating particles such as muons survive to ground. Muons are accompanied by an EM component produced mainly by muon decay in flight and muon interactions such as bremsstrahlung, pair production and nuclear interactions, which amount to $\sim 20\%$ of the muonic component [1]. This is the so-called electromagnetic "halo".

The Surface Detector Array (SD) of the Pierre Auger Observatory [2] is well suited to detect very inclined showers at energies above about $5 \times 10^{18}$ eV, with high efficiency and unprecedented statistical accuracy. The cosmic ray energy spectrum obtained with inclined events is given in these proceedings [3].

The distribution of the detector signals produced by shower particles is used to estimate shower observables such as the primary energy. The specific characteristics of inclined showers, such as the absorption of the EM component and the deviations suffered by muons in the geomagnetic field, entail that their analysis requires a different approach from the standard one for showers of $\theta < 60°$. The study of the signal distributions of the electromagnetic and muonic components at ground level becomes essential in the reconstruction [3], [4] and analysis of events at large angles.

In this work we have performed a comprehensive characterisation of the electromagnetic component with respect to the well-known behaviour of the muonic component. We have studied the ratio of the EM to muonic contributions to the signal in the water-Cherenkov detector as a function of several parameters. We have examined the effect of the shower evolution, shower geometry and geomagnetic field on the ratio. The dependences of this ratio on the primary energy, mass composition and hadronic model assumed in the simulations are addressed. The resulting parameterisations are used for the reconstruction of inclined events measured with the SD of the Pierre Auger Observatory [3].

The study described here is based on Monte Carlo simulations. A library of proton and iron-induced showers with energies from $10^{18}$ to $10^{20}$ eV, zenith angles between $60°$ and $88°$ and random azimuthal angle were generated with AIRES 2.6.0 [5] and the hadronic interaction models QGSJET01 [6] and Sibyll 2.1 [7]. The showers were simulated with and without geomagnetic field at the site of the SD of the Pierre Auger Observatory. The detector response is calculated here using a simple method based on parameterisations of the detector response to the passage of shower particles.

## II. THE RATIO OF ELECTROMAGNETIC TO MUONIC DETECTOR SIGNALS

The electromagnetic and muonic particle components have a characteristic behaviour with distance to the shower axis, shower zenith angle and azimuth angle ($\zeta$) of the detector position with respect to the incoming shower direction projected onto the plane transverse to the shower axis (shower plane). Also the different contributions to the electromagnetic component differ from each other as shown below. This is reflected by the ratio of the EM to muonic contributions to the detector signal

$$R_{\mathrm{EM}/\mu} = S_{\mathrm{EM}}/S_{\mu} \qquad (1)$$

In Fig. 1, we show the average signal distributions of the EM and muonic components (left panel) and their corresponding ratio $R_{\mathrm{EM}/\mu}$ (right panel) as a function of the distance to the core $r$ for different $\theta$. Near the core, the ratio decreases with zenith angle from $\theta = 60°$ to $\sim 70°$ because the remnant of the EM shower due to cascading processes ($\pi^0$ decay) is increasingly absorbed, until it practically disappears at $\theta \sim 70°$. Then the ratio increases again with $\theta$, mainly due to muon hard interaction processes (bremsstrahlung, pair production and nuclear interactions) that are expected to dominate near the core in very inclined showers. Far from the core the lateral distribution of the ratio tends to flatten due to the dominant contribution of the EM halo produced by muon decay in flight. The larger the zenith angle, the ratio levels off closer to shower core. The slight increase of the ratio for $\theta \lesssim 68°$ and far from the core ($r \gtrsim 2$ km ) is attributed to the combination of two effects, one is that the number of low energy muons decreases more rapidly at large distances because they decay before reaching the ground, and only energetic muons survive, and on the other hand the presence of the contribution to the EM component due to $\pi^0$ decay, particularly in the early region of the shower (the portion of the shower front that hits the ground before the shower axis).

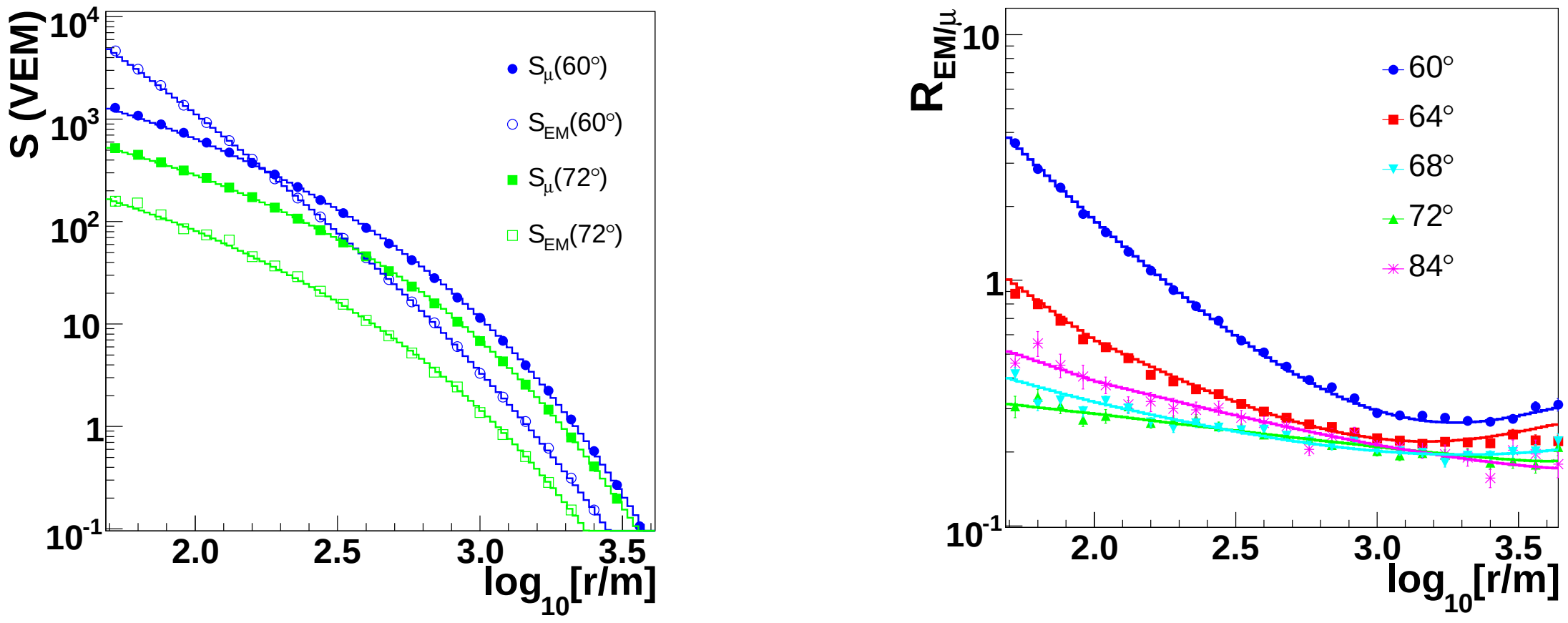

Fig. 1. Left plane: Lateral distribution of the electromagnetic and muonic contributions to the signal in the shower plane. Right panel: The ratio of the electromagnetic to muonic contributions to the detector signal as a function of the distance from the shower axis. Simulations were performed for 10 EeV proton showers at different zenith angles and in absence of geomagnetic field.

### A. Azimuthal asymmetry of the ratio $R_{\mathrm{EM}/\mu}$

There is an azimuthal asymmetry in the ratio of the EM to muonic contributions due to the combination of the geometrical and shower evolution effects [8]. As illustrated in the left panel of Fig. 2 shower particles do not travel parallel to the shower axis in general and therefore they cross different amounts of atmosphere depending on $\zeta$. In particular, particles arrive at ground in the early region of the shower ($\zeta = 0°$) with a smaller local zenith angle than those in the late region ($\zeta = 180°$). This is essentially the basis for the geometrical effect. In inclined showers, the asymmetry induced by the geometrical effect is typically small and the main source of azimuthal asymmetry is the shower evolution effect which can be understood as follows. Particles at the same distance from the shower axis in the shower plane, but arriving with different $\zeta$, travel along different paths and belong to different stages in the evolution of the shower. The importance of this effect depends on the depth-dependent evolution of the lateral particle distribution and on the attenuation of the total number of particles. The asymmetry induced by the shower evolution affects more the remnant of the EM shower than the muonic component or its associated EM halo. As a consequence, the shower evolution is expected to induce a negligible asymmetry in the ratio in showers with $\theta \gtrsim 70°$, because the EM remnant is practically suppressed, and the EM halo approximately has the same asymmetry than the muonic component.

To study further the azimutal dependence of the asymmetry we divide the shower plane in $\zeta$ bins, and we calculate the lateral distributions of the ratio in each bin for a fixed zenith angle: $R_{\mathrm{EM}/\mu}(r,\theta,\zeta)$, and we compare these distributions to the distribution obtained averaging over $\zeta$: $\langle R_{em/\mu}\rangle(r,\theta)$. For this purpose we define the asymmetry parameter $\Delta_\zeta$ as

$$R_{\mathrm{EM}/\mu}(r,\theta,\zeta) = \langle R_{\mathrm{EM}/\mu}\rangle(r,\theta) \times (1+\Delta_\zeta) \qquad (2)$$

In Fig. 2, we show the lateral distribution of $R_{\mathrm{EM}/\mu}$ in different $\zeta$ bins compared to the mean value (middle panel) and their corresponding asymmetry parameter $\Delta_\zeta$ (right panel) for showers at $\theta = 60°$. $|\Delta_\zeta|$ increases with distance to the core and it is larger in the early region than in the late region as expected. Moreover, $|\Delta_\zeta|$ decreases as the zenith angle increases for the reasons explained above, becoming negligible for $\theta > 68°$. This plot illustrates the importance of accounting for the asymmetry in the ratio when dealing with inclined showers with $60° < \theta < 70°$.

### B. Geomagnetic field effect on $R_{\mathrm{EM}/\mu}$

Muons in inclined showers travel along sufficiently long paths in the atmosphere to be affected by the Earth's magnetic field (GF). Positive and negative muons are deviated in opposite directions and as a consequence the muonic patterns in the shower plane are distorted in elliptical or even 2-lobed patterns [9], [10]. This effect on the muonic distributions is only significant for $\theta \geq 75°$. At these angles, the dominant contribution to the EM signal at ground is due to the EM halo, which inherits the muon spatial distribution and is proportional to the muonic signal distribution. For this reason, the ratio of the EM to muonic signals maintains the symmetry in the azimuthal angle $\zeta$. However, the GF increases the $\langle R_{\mathrm{EM}/\mu}\rangle$ with respect to the value in its absence. The effect depends on the shower zenith ($\theta$) and azimuth ($\phi$) angles, and is more important near the core. After studying all these dependences, we have concluded that the effect on the ratio is important for showers at $\theta \gtrsim 86°$. It should be noted that the rate of events at such high zenith angles detected at ground level is small due to the reduced solid angle and the $\cos\theta$ factor needed to project the array area onto the shower plane. Very inclined events are also subject to other uncertainties [3] and we therefore choose to ignore them at this stage without losing much on statistical grounds.

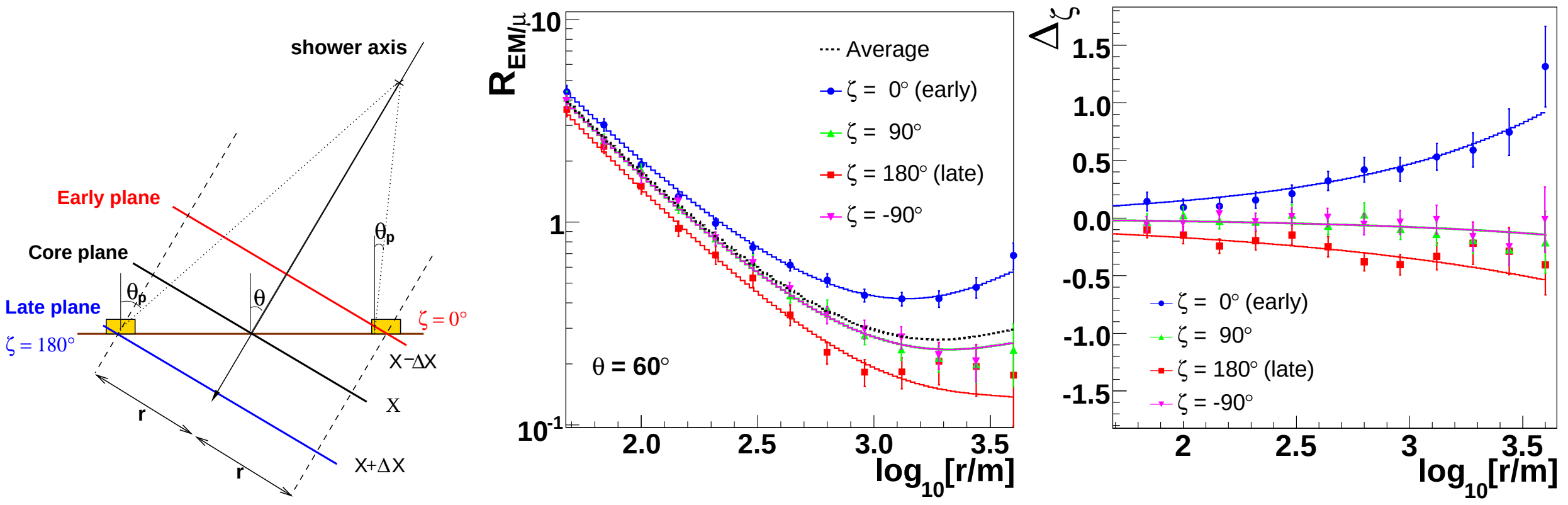

Fig. 2. Azimuthal asymmetry in the ratio $R_{\mathrm{EM}/\mu}$. Left panel: Schematic picture of an inclined shower reaching the ground. Middle panel: The ratio $R_{\mathrm{EM}/\mu}$ as a function of the distance from the shower axis in the shower plane in different bins of $\zeta$ for 10 EeV proton showers with $\theta = 60°$. Right panel: Asymmetry of the lateral distribution of the ratio $R_{\mathrm{EM}/\mu}$ in different $\zeta$ bins. The size of the bins is $\Delta\zeta = 30°$ centered at $\zeta$.

### C. *Systematic uncertainties*

The lateral distributions of the electromagnetic signal due to cascading processes and muonic signal exhibit a different behaviour as a function of the energy and of the depth of the shower maximum, while the contribution to the EM signal due to muon decay in flight mimics the energy dependence of the muonic one. Combining all the results, we expect $R_{\mathrm{EM}/\mu}$ to have a different behaviour depending on whether the EM remnant or the EM halo contributes more to the total signal. We study the energy dependence of $R_{\mathrm{EM}/\mu}$ performing the relative difference $\Delta_E$ between the ratio at a given energy with respect to that obtained for 10 EeV proton-induced showers, $\langle R_{\mathrm{EM}/\mu}\rangle$:

$$\Delta_E = \frac{R_{\mathrm{EM}/\mu}(E) - \langle R_{\mathrm{EM}/\mu}\rangle(10\mathrm{EeV})}{\langle R_{\mathrm{EM}/\mu}\rangle(10\mathrm{EeV})} \qquad (3)$$

The dependence of $\Delta_E$ on the zenith angle and distance from the shower axis is studied as in the example of Fig. 3 (left panel), where we plot $\Delta_E$ in different bins of r, as a function of the zenith angle for 1 EeV proton showers. We find that either for $\theta \gtrsim 68°$ at all the distances to the shower core or for distances beyond 1 km at all the zenith angles the ratio $R_{\mathrm{EM}/\mu}$ remains constant at the same level with energy because only the EM halo contributes to the EM signal. Otherwise, there is a dependence on energy that increases as the distance to the shower axis decreases, and therefore the dependences must be taken into account as systematic uncertainties. We obtain the same general result studying $\Delta_E$ for other shower energies.

At present, the chemical composition of the cosmic rays at the highest energies ($> 1$ EeV) remains unknown. For this reason we have studied the dependence of the ratio on the mass of the primary particle initiating the shower accounting for protons and iron nuclei in our simulations. Following the same procedure as in the case of the energy, we calculate the relative difference $\Delta_{\mathrm{mass}}$ between the ratio in iron-induced showers at 10 EeV with respect to that obtained for 10 EeV proton shower simulations:

$$\Delta_{\mathrm{mass}} = \frac{R_{\mathrm{EM}/\mu}(\mathrm{Fe}) - \langle R_{\mathrm{EM}/\mu}\rangle(\mathrm{p})}{\langle R_{\mathrm{EM}/\mu}\rangle(\mathrm{p})} \qquad (4)$$

For reasons very similar to those that explain the energy dependence studied before, we conclude that either for $\theta \gtrsim 68°$ at all the distances to the shower core or for distances beyond 1 km at all $\theta$ the ratio $R_{\mathrm{EM}/\mu}$ remains constant at the same level with primary mass as shown in Fig. 3 (middle panel).

At the highest energies, there is lack of knowledge about the hadronic interactions which determine the shower development of MC simulations [11]. This fact leads to discrepancies between the different hadronic models on predictions such as the densities of the EM and muonic components at ground.

In this work, we compare two high energy interaction models widely used in cosmic ray physics: QGSJET01

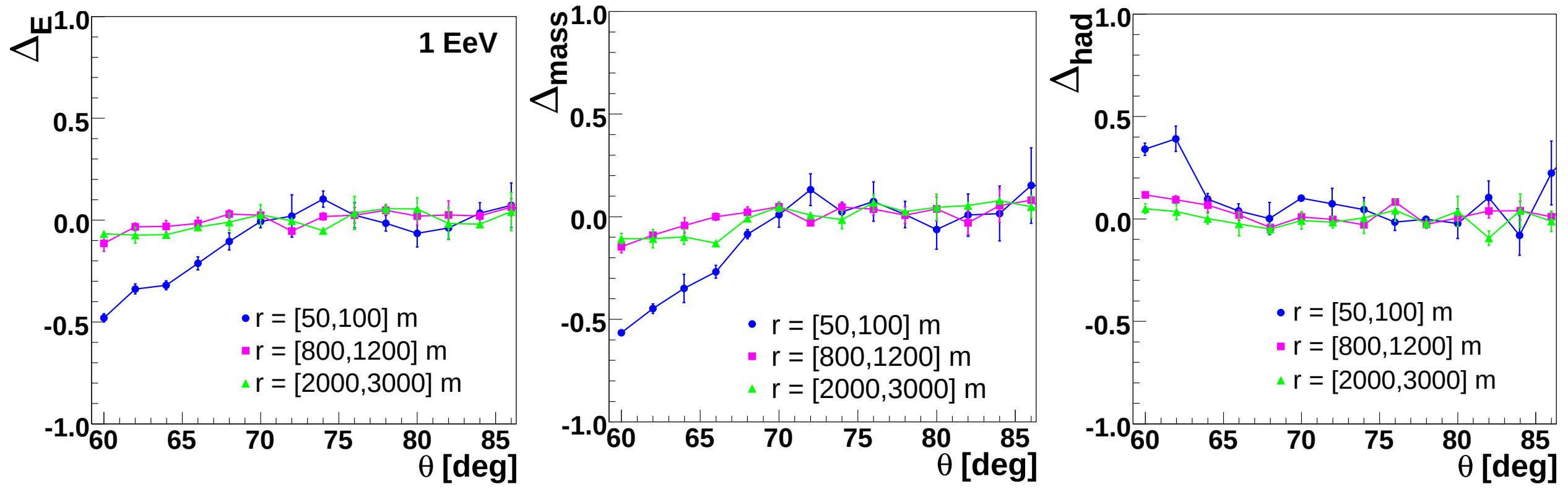

Fig. 3. Left plane: The relative difference $\Delta_{\rm E}$ between the ratio $R_{\rm EM/\mu}$ obtained in 1 EeV proton-induced showers with respect to the reference ratio $\langle R_{\rm EM/\mu}\rangle$ obtained in 10 EeV proton-induced showers simulated with QGSJET01 (Eq. 3). Middle panel: The relative difference $\Delta_{\rm mass}$ between the ratio $R_{\rm EM/\mu}$ obtained in 10 EeV iron-induced showers simulations with respect to $\langle R_{\rm EM/\mu}\rangle$ (Eq. 4). Right panel: The relative difference $\Delta_{\rm had}$ between the ratio $R_{\rm EM/\mu}$ obtained in 10 EeV proton-induced showers simulated with Sibyll 2.1 with respect to $\langle R_{\rm EM/\mu}\rangle$ (Eq. 5). The relative differences are shown as a function of the shower zenith angle in different bins of distance to the shower axis r.

and Sibyll 2.1. For proton primaries at 10 EeV, the QGSJET model predicts showers that on average develop higher in the atmosphere and have $40\%$ more muons than showers simulated with Sibyll.

We calculate the relative difference $\Delta_{\rm had}$ between the ratio for 10 EeV proton showers simulated with Sibyll 2.1 with respect that obtained in showers simulated with QGSJET01:

$$\Delta_{\rm had} = \frac{R_{\rm EM/\mu}({\rm Sibyll}) - \langle R_{\rm EM/\mu}\rangle({\rm QGSJET})}{\langle R_{\rm EM/\mu}\rangle({\rm QGSJET})} \quad (5)$$

In Fig. 3 (right panel) we show $\Delta_{\rm had}$ as a function of the zenith angle in different bins of $r$. The differences between both models are more apparent near the shower axis as expected from the dominance of the EM component due to cascading processes near the core. We obtain a similar result to the case of energy and mass dependences, which is that either for $\theta \gtrsim 64°$ at all the distances to the shower axis or for distances beyond 1 km at all zenith angles the ratio $R_{\rm EM/\mu}$ remains constant at the same level independently of the model used.

## III. Conclusions

We have characterised the signal distributions of the electromagnetic and muonic components of inclined showers at the ground level on the shower plane [12]. We have accounted for the different sources of azimuthal asymmetry and the effect of the geomagnetic field. As a result, we have obtained a parameterisation of the ratio $S_{\rm EM}/S_\mu$ as a function of the shower zenith angle and the detector position that is used in the reconstruction of inclined events measured with the Surface Detector Array of the Pierre Auger Observatory.

We have studied the dependence of this ratio with the primary energy, mass composition and the hadronic interaction model used in the simulations. The general result is that either for zenith angles exceeding $\theta \gtrsim 68°$ or for distances to the shower core beyond 1 km at all the zenith angles $> 60°$, the ratio remains constant because only the electromagnetic halo contributes to the EM signal. Otherwise, the dependences are important and must be taken into account as systematic uncertainties within the event reconstruction.

# Acknowledgements

The successful installation and commissioning of the Pierre Auger Observatory would not have been possible without the strong commitment and effort from the technical and administrative staff in Malargüe.

We are very grateful to the following agencies and organizations for financial support:

Comisión Nacional de Energía Atómica, Fundación Antorchas, Gobierno De La Provincia de Mendoza, Municipalidad de Malargüe, NDM Holdings and Valle Las Leñas, in gratitude for their continuing cooperation over land access, Argentina; the Australian Research Council; Conselho Nacional de Desenvolvimento Científico e Tecnológico (CNPq), Financiadora de Estudos e Projetos (FINEP), Fundação de Amparo à Pesquisa do Estado de Rio de Janeiro (FAPERJ), Fundação de Amparo à Pesquisa do Estado de São Paulo (FAPESP), Ministério de Ciência e Tecnologia (MCT), Brazil; AVCR AV0Z10100502 and AV0Z10100522, GAAV KJB300100801 and KJB100100904, MSMT-CR LA08016, LC527, 1M06002, and MSM0021620859, Czech Republic; Centre de Calcul IN2P3/CNRS, Centre National de la Recherche Scientifique (CNRS), Conseil Régional Ile-de-France, Département Physique Nucléaire et Corpusculaire (PNC-IN2P3/CNRS), Département Sciences de l'Univers (SDU-INSU/CNRS), France; Bundesministerium für Bildung und Forschung (BMBF), Deutsche Forschungsgemeinschaft (DFG), Finanzministerium Baden-Württemberg, Helmholtz-Gemeinschaft Deutscher Forschungszentren (HGF), Ministerium für Wissenschaft und Forschung, Nordrhein-Westfalen, Ministerium für Wissenschaft, Forschung und Kunst, Baden-Württemberg, Germany; Istituto Nazionale di Fisica Nucleare (INFN), Ministero dell'Istruzione, dell'Università e della Ricerca (MIUR), Italy; Consejo Nacional de Ciencia y Tecnología (CONACYT), Mexico; Ministerie van Onderwijs, Cultuur en Wetenschap, Nederlandse Organisatie voor Wetenschappelijk Onderzoek (NWO), Stichting voor Fundamenteel Onderzoek der Materie (FOM), Netherlands; Ministry of Science and Higher Education, Grant Nos. 1 P03 D 014 30, N202 090 31/0623, and PAP/218/2006, Poland; Fundação para a Ciência e a Tecnologia, Portugal; Ministry for Higher Education, Science, and Technology, Slovenian Research Agency, Slovenia; Comunidad de Madrid, Consejería de Educación de la Comunidad de Castilla La Mancha, FEDER funds, Ministerio de Ciencia e Innovación, Xunta de Galicia, Spain; Science and Technology Facilities Council, United Kingdom; Department of Energy, Contract No. DE-AC02-07CH11359, National Science Foundation, Grant No. 0450696, The Grainger Foundation USA; ALFA-EC / HELEN, European Union 6th Framework Program, Grant No. MEIF-CT-2005-025057, European Union 7th Framework Program, Grant No. PIEF-GA-2008-220240, and UNESCO.